# Designing and evaluating advanced adaptive randomised clinical trials: a practical guide

Anders Granholm[1,*], Aksel Karl Georg Jensen[1,2], Theis Lange[2],
Anders Perner[1,3], Morten Hylander Møller[1,3], and Benjamin Skov Kaas-Hansen[1,2]

[1] Department of Intensive Care 4131, Copenhagen University Hospital – Rigshospitalet, Copenhagen, Denmark
[2] Section of Biostatistics, Department of Public Health, University of Copenhagen, Copenhagen, Denmark
[3] Department of Clinical Medicine, Faculty of Health and Medical Sciences, University of Copenhagen, Copenhagen, Denmark

**\* Correspondence:**
Anders Granholm, MD, PhD
Department of Intensive Care 4131, Copenhagen University Hospital – Rigshospitalet
DK-2100 Copenhagen, Denmark
Email: anders.granholm@regionh.dk

**Version date:** 2025-01-15

**Word count main text:** 6794
**Word count abstract:** 192

**Key words:** randomised clinical trial; trial design; adaptive trials; simulation; randomisation; response-adaptive randomisation.

# Abstract

**Background**
Advanced adaptive randomised clinical trials are increasingly used. Compared to their conventional counterparts, their flexibility may make them more efficient, increase the probability of obtaining conclusive results without larger samples than necessary, and increase the probability that individual participants are allocated to more promising interventions. However, limited guidance is available on designing and evaluating the performance of advanced adaptive trials.

**Methods**
We summarise the methodological considerations and provide practical guidance on the entire workflow of planning and evaluating advanced adaptive trials using adaptive stopping, adaptive arm dropping, and response-adaptive randomisation within a Bayesian statistical framework.

**Results**
This comprehensive practical guide covers the key methodological decisions for advanced adaptive trials and their specification and evaluation using statistical simulation. These considerations include interventions and common control use; outcome type and generation; analysis timing and outcome-data lag; allocation rules; analysis model; adaptation rules for stopping and arm dropping; clinical scenarios assessed; performance metrics; calibration; sensitivity analyses; and reporting. The considerations are covered in the context of realistic examples, along with simulation code using the *adaptr* R package.

**Conclusions**
This practical guide will help clinical trialists, methodologists, and biostatisticians design and evaluate advanced adaptive trials.

# Background

Randomised clinical trials (RCTs) constitute the most rigorous research design for unbiased comparative effectiveness estimates of healthcare interventions [1]. However, conventional RCTs are limited by their inflexibility [1, 2]. Most conventional RCTs use a fixed maximum sample size with no or few interim analyses. Sample size calculations often rely on over-optimistic assumptions [3–6], which pose the risk that trials will be unable to provide firm conclusions about smaller, yet clinically relevant effects [1, 2]. Unfortunately, such results may incorrectly be interpreted as no difference between the interventions [7–9]. Also, RCTs may run longer than necessary, which may harm trial participants in inferior arms, delay implementation of superior interventions, delay de-implementation of inferior interventions, and result in wasted research funding [1, 2, 10, 11]. Finally, except when stopped early, results are typically not used until the RCT has enrolled the planned maximum sample size, precluding continuous learning during the course of the trial [10, 12].

Adaptive trials use results from adaptive (interim) analyses to modify some aspects of the trial before completion, *without* undermining the integrity and validity of the trial [13]. The most common adaptive trials are *group sequential designs* [13], but they typically use only few interim analyses with strict thresholds for early stopping [2]. There is an increased use of advanced adaptive trials (including adaptive platform trials) [10, 14, 15], which often use multiple adaptive features and involve many more adaptive analyses than conventional group sequential trials [1, 2, 13].

Advanced adaptive trials may be stopped entirely, or specific arms dropped, for several reasons – e.g., inferiority/superiority, practical equivalence, futility, or at a pre-specified maximum sample size – shortening the time required to reach valid and conclusive results [2, 10]. Adaptive arm dropping prioritises allocation to more promising interventions and increases power for the remaining comparisons when inferior arms are dropped early in trials with >2 arms [2, 10]. Response-adaptive randomisation increases allocation to arms more likely to be superior *before* the evidence is sufficient for overall termination of the trial [2, 10]. Both features can increase the probability of beneficial outcomes for randomised participants as the trial progresses [2, 10].

The increased flexibility, however, comes at a cost: everything else being equal, more adaptive analyses increase the risk of stopping or adapting due to chance findings. Further, while response-adaptive randomisation may increase the chances of better outcomes for individual participants, it may also increase the overall required sample sizes in some cases and, unless adequately restricted, adaptations to random fluctuations may substantially impair trial performance either on average or in the worst case scenario [2, 16–23]. As such, there have been ethical arguments both in favour of and against increased adaptation, especially response-adaptive randomisation [20–24].

Consequently, it is paramount that the performance of adaptive trial designs are carefully evaluated prior to initiation [1, 2, 13], and this is generally required by the competent authorities [25, 26]. In contrast to conventional non-adaptive and less advanced adaptive trials such as two-armed group sequential trials [13], this cannot be achieved with simple closed-form sample size calculations and readily available methods for defining stopping rules. Instead, statistical simulation is required [2, 25–28]. Planning advanced adaptive trials may hence seem daunting due to the more comprehensive processes that require specific methodological and statistical competences.

To alleviate this, we provide a practical guide covering the entire process of planning and evaluating an advanced Bayesian adaptive trial using the *adaptr* [27] R package. We provide guidance on how to use simulations to evaluate and compare advanced adaptive designs and examples of the code required to do so.

# Overview

## Scope and target audience

The key phases in an RCT are 1) identification of the clinical problem and formulation of the research question; 2) trial design; 3) trial conduct; and 4) analysis and reporting. The scope of this manuscript is to provide an example-based practical guide on key steps necessary for designing advanced adaptive trials, including use of simulations for assessing their performance (e.g., expected sample sizes, type 1 error rates, power, etc.), corresponding to those parts of the third phase that are specific to advanced adaptive trials. As such, methodological considerations relevant regardless of the adaptations (e.g., setting, number of centres, use of blinding, procedures for inclusion and follow-up, etc.) are not covered here. We focus on phase 3 or 4 comparative effectiveness trials with adaptive stopping, arm dropping, and/or response-adaptive randomisation using a Bayesian framework, as is common in advanced adaptive trials [10, 14]. The target audience is clinical trialists, methodologists, and biostatisticians with previous experience on RCT planning and conduct and, ideally, basic knowledge of Bayesian statistics, adaptive trial designs, and the R statistical software. Additional guidance and information on other aspects of adaptive trial planning and conduct can be found elsewhere [2, 10, 13] (including the *PANDA* [panda.shef.ac.uk] and *CTTI* [ctti-clinicaltrials.org/our-work/novel-clinical-trial-designs] repositories).

## Bayesian statistics and advanced adaptive trials

Within a Bayesian statistical framework, uncertainty is expressed using probability distributions [29]. In brief, results are expressed as a *posterior* probability distribution that is a weighted compromise between a *prior* probability distribution reflecting the belief before obtaining the new data, and the observed data expressed via a likelihood function [29], as illustrated in **Fig. 1**.

**Fig. 1**

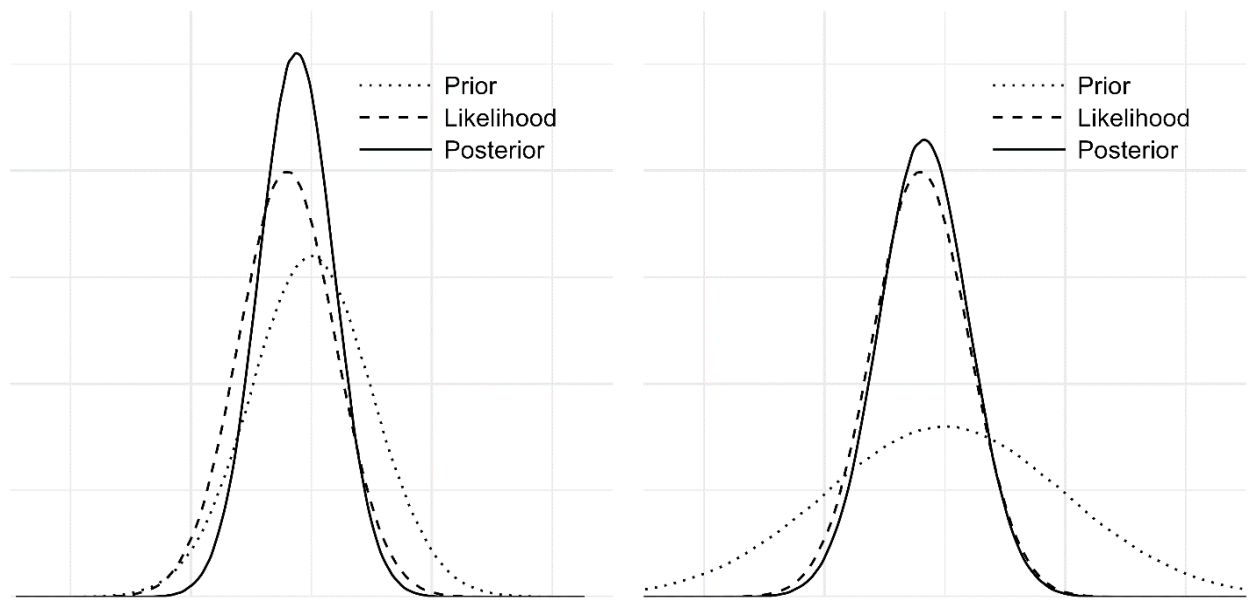

Figure legend: Illustration of probabilities in Bayesian analyses. Horizontal axes represent specific values on the outcome scale (on a fictive, unitless scale in this example), while vertical axes represent densities (with higher values being more probable). Posterior probability distributions (full lines) combine prior probability distributions (dotted lines) with the obtained data through a likelihood function (dashed lines), with the posterior probability being a weighted compromise between the prior and the data. The left subplot illustrates a relatively informative (narrow) prior distribution with the resulting posterior centred between the prior and the likelihood, but more precise (i.e., narrower) than both. The right subplot illustrates a less informative (wider) prior, where the posterior largely overlaps with the likelihood and is only slightly more precise than the likelihood itself. The data and likelihood functions are identical in both subplots.

The Bayesian statistical approach is well-suited for advanced adaptive trial designs as the implementation and evaluation of the adaptation rules is relatively simple once the posterior distributions are available. Of note, many Bayesian adaptive trials may technically be considered hybrid Bayesian-frequentist, as Bayesian analogues of inherently frequentist concepts such as type 1 error rates and power are evaluated using long-run frequencies from statistical simulations [2, 30]. While the importance of *always* tightly controlling these metrics has been discussed [31, 32], the competent authorities typically require this for late-phase trials [25, 26], and so may funders and ethical committees [33]. It is thus usually recommended and done [16, 34], as it is otherwise difficult to ensure that the adaptive features do not challenge the validity of the trial [13].

### Contents
The rest of the paper summarises the relevant key methodological considerations and covers how trial designs are specified and evaluated with simulations using the *adaptr* [27] R package, including sensitivity analyses, and reporting of results. Finally, a brief discussion is provided covering limitations with the described approach and the package.

The *adaptr* [27] package simulates adaptive (multi-arm, multi-stage) RCTs using adaptive stopping and arm dropping for superiority, inferiority, practical equivalence, and/or futility, as well as fixed and/or response-adaptive randomisation. We used *adaptr* v1.4.0 with R v4.4.1; complete details on simulation options, arguments, and additional functions in *adaptr*, including visualisation functions, can be found in the package documentation (inceptdk.github.io/adaptr).

## Methodological choices and simulation-based evaluation

The key methodological considerations that we cover are presented in **Fig. 2**; importantly, methodological decisions interact, and as such the development and evaluation of an advanced adaptive trial design will typically be an iterative process [2].

**Fig. 2**

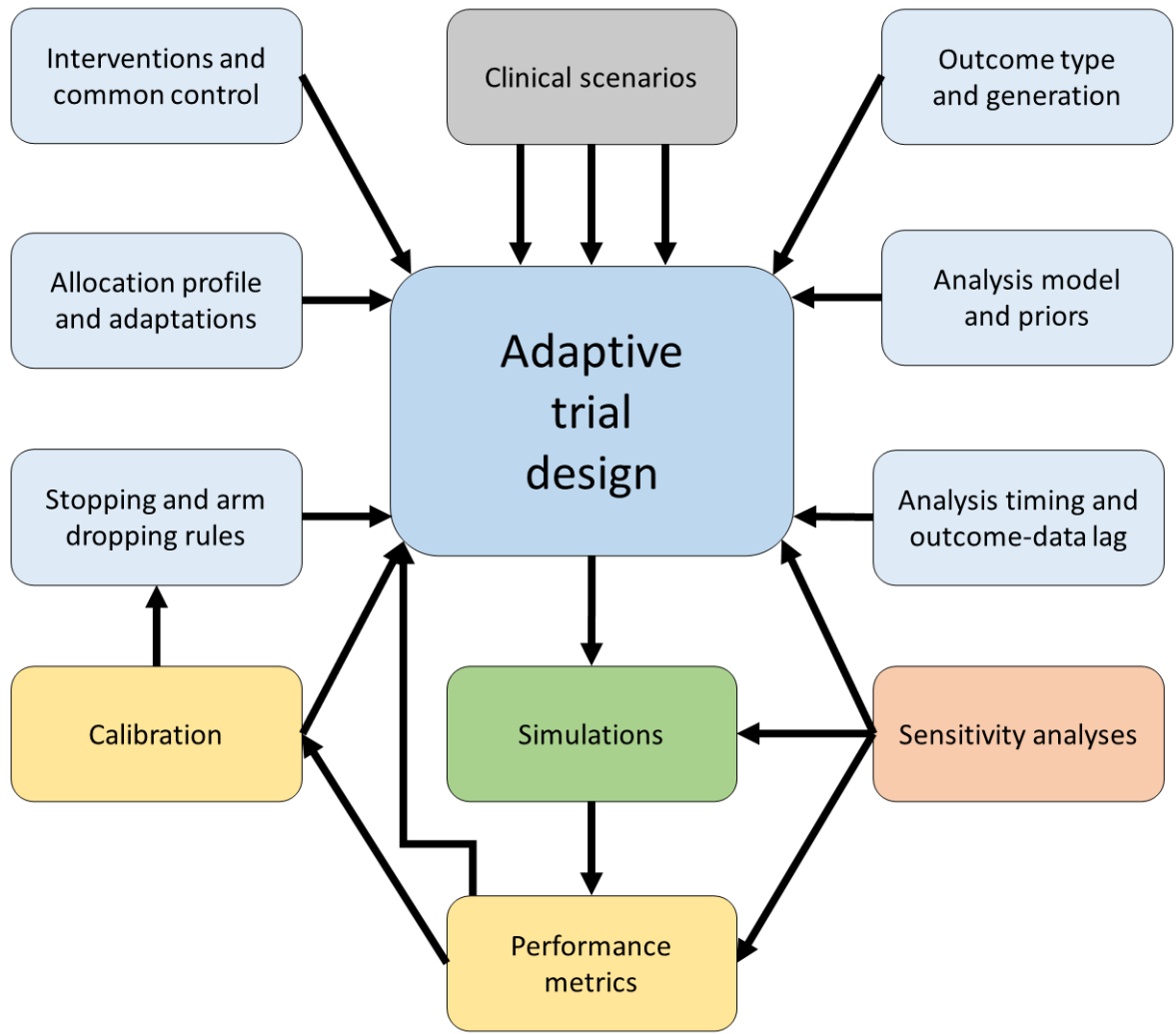

Figure legend: Overview of the process of designing and evaluating advanced adaptive trials, with focus on methodological decisions related to the trial design, the simulation process, measurement of performance metrics (e.g., expected sample size, type 1 error rates, power, etc.), calibration, iterative revisions of the trial design, and finally, the use of sensitivity analyses to assess the influence of different design choices and assumptions on performance. Light blue boxes cover essential design choices that may be varied, but where the final design will only be based on a single choice for each option; *clinical scenarios* (grey box) are similarly essential, but generally, multiple scenarios will be evaluated for each single combination of all other design choices. Simulations (green box) are used to evaluate trial designs (i.e., calculate performance metrics) across different clinical scenarios and to optionally *calibrate* stopping rules to obtain acceptable values for one or more performance metrics (both in yellow boxes). Finally, sensitivity analyses of design choices and uncontrollable assumptions (orange box) will typically be used to evaluate the implications of different design choices and other plausible assumptions on performance metrics through simulations. Figure inspired by a figure in a previous article by our group [2]; additional guidance on the methodological choices may also be found in that article.

The following sections describes the key methodological choices in general along with details on their specific implementation in *adaptr* and an example including code, mostly in that order. The example trial uses three interventional arms without a common control arm, an undesirable binary outcome, restricted response-adaptive randomisation, stopping and arm dropping for superiority/inferiority and practical equivalence, and a maximum sample size of 10,000 participants. Complete code and all outputs from the primary example are included in **appendix 1**.

**Setup**

First, load *adaptr*, set up a cluster for parallel computation for faster simulations, and define where results are saved:

```
library(adaptr)
setup_cluster(10) # Number of cores for parallel computation
dir_out <- "<PATH>/" # Replace with an actual, permanent path
```

**Trial design**

In *adaptr*, trial designs including outcome generation and analysis models are specified via the `setup_trial()` function in the general case. For binary, binomially distributed outcomes, `setup_trial_binom()` can be used and has very weak, flat priors (*Beta(alpha = 1, beta = 1)* priors, which corresponds to two randomised participants, one with the outcome and one without [35]). For continuous, normally distributed outcomes, `setup_trial_norm()` can be used and uses no prior information. Our primary example uses `setup_trial_binom()`. In the following, we only include the arguments specifying the options discussed in each section (with omitted parts of the code marked with `...`), with the complete trial design specified at the end of this section. We explicitly specify certain key arguments for clarity, even when identical to the defaults.

*Interventions and use of common control*

The initial trial interventions (arms) must be specified, and a *common control* arm may be specified if relevant; use of a common control arm will influence trial behaviour for multi-arm designs. Without a common control arm, all adaptive decisions will be based on the probabilities of each arm being overall best or of all arms being practically equivalent. With a common control arm, all other arms will be compared *pairwise* against it with all stopping/arm dropping decisions based on probabilities from the pairwise comparisons (section *"Adaptation rules for stopping and arm dropping"*). If a non-control arm is superior, the current common control is dropped, and the superior arm promoted to the new control arm. This is followed by immediate pairwise comparisons against the remaining non-control arms, before inclusion of additional simulated participants. If multiple non-control arms are superior, the one with the highest probability of being overall best is chosen. Even when one arm represents *standard of care*, we advise that designs both with and without a formal common control are assessed due to the influence of this decision on a trial design's performance.

Here, we include three arms without a formal common control:

```
setup_trial_binom(
  arms = c("Arm A", "Arm B", "Arm C"),
  control = NULL,
  ...
)
```

*Outcome type and generation*
A single *guiding* outcome is simulated and used for all adaptations. Typically, this will be the primary trial outcome, but an *intermediary* or *surrogate* outcome, e.g., the same outcome after a shorter follow-up period or another outcome that is highly correlated with the primary outcome may be chosen based on careful considerations [2, 13, 36]. To generate outcome data, the outcome distribution must be defined. Advanced adaptive trial designs should be assessed under multiple different clinical *scenarios*, i.e., different sets of assumed outcome distributions in each arm, which must be specified as part of separate, but otherwise identical, trial specifications (further details in section *"Clinical scenarios"*). We recommend initially specifying a scenario without differences present and using the clinically most plausible outcome distribution as the reference.

In *adaptr*, outcomes must be numerical, *even* if they correspond to, e.g., binary or ordinal outcomes. Further, we must specify whether higher or lower values are desirable. The example uses an undesirable, binary, binomially distributed outcome – e.g., mortality – encoded to reflect the common encoding of mortality: 0 denotes survival (no event) while 1 denotes death (event). Here, the (assumed) true event probabilities are 25% in all arms, reflecting no between-arm differences:

```
setup_trial_binom(
  ...
  true_ys = c(0.25, 0.25, 0.25),
  highest_is_best = FALSE,
  ...
)
```

*Analysis timing and outcome-data lag*
The number of participants analysed at the time of each adaptive analysis must be specified along with a adequately large *burn-in*: an initial period where no adaptations occur before a sufficient number of participants are included. This prevents adaptations to early, random fluctuations [2, 18, 24, 36] and avoids stopping trials or dropping arms with samples so small that results may be considered unreliable or that precision of effects on important outcomes may be considered too low. The timing of subsequent adaptive analyses should consider both the maximum total number of analyses, the maximum allowed sample size, and the expected inclusion rates. While more analyses mean that stricter stopping thresholds may be required (section *"Adaptation rules for stopping and arm dropping"*), a lower number of analyses will limit the potential benefits of the adaptive design [2, 24].

Importantly, the outcome-data lag and the expected inclusion rate should be considered, as both will affect the efficiency and reliability of adaptations and ultimately the performance metrics [37]. Outcome-data lag is the outcome follow-up duration plus the expected time required to obtain, clean, and validate data before analyses can be conducted [37]. The expected inclusion rate can be constant over time or change, e.g., if the number of active trial sites is expected to change. Longer outcome-data lags or higher inclusion rates mean that the proportion of randomised participants with data available at the time of each adaptive analysis will be lower, increasing the risk that results will change (in direction or magnitude) at the final analysis conducted after stopping enrolment and completing follow-up for all randomised participants [37, 38]. It has been suggested that the ratio between the outcome follow-up duration and the expected inclusion period should be <0.25 for adaptive trials to be beneficial [39]. In previous trials by our group, inclusion rates were mostly constant after initiation of all participating trial sites [40–44], but this will vary between trials. Some advanced adaptive trials [45] are planned *without* a maximum sample size; these can be simulated using *adaptr* by setting an implausibly high maximum sample size and ensuring that a stopping rule will always be triggered before the specified sample size limit (section *"Adaptation rules for stopping and arm dropping"*). Analysis timing is typically based on the number of participants that have completed their outcome-data lag period and can be included in the analysis and not on the total number of participants randomised.

Below, we specify that the first analysis will be conducted after 500 participants have available data, with subsequent analyses after each 250 additional participants up to a maximum sample size of 10,000 participants. The example assumes a constant lag of 200 participants:

```
setup_trial_binom(
  ...
  # Number of participants with data available and included in each analysis
  data_looks = seq(from = 500, to = 10000, by = 250),
  # Number of participants randomised at each analysis
  randomised_at_looks = c(seq(from = 700, to = 9950, by = 250), 10000),
  # Note: the maximum number in both arguments should be equal
  ...
)
```

*Allocation profiles*
The initial allocation profile and subsequent use of fixed allocation, response-adaptive randomisation, or combinations must be specified, including any restrictions. Although response-adaptive randomisation increases the probability of allocating more participants to more promising interventions [2, 16, 23], there are ethical, practical, and logistical arguments both in favour and against its use [20–24, 46, 47].

Response-adaptive randomisation affects trial performance (section *"Performance metrics"*) differently depending on the number of arms, whether a common control arm is used, and whether between-arm differences are present [2, 17–21, 23, 37, 48, 49]. Previous results indicate that fixed allocation or relatively restricted response-adaptive randomisation may be preferable in two-arm trials; both fixed and relatively restricted response-adaptive allocation may perform well in trials with >2 arms with no common control; and a relatively higher, fixed allocation probability to the control arm and response-adaptive randomisation to non-control arms may be preferable in trials with >2 arms and a common control [2, 33]. Importantly, response-adaptive randomisation may improve certain performance metrics while worsening others: higher probabilities of desirable outcomes for individual participants, for example, may on average require larger samples [2, 17–21, 23, 37, 48, 49]. Further, even when response-adaptive randomisation on average improves performance, it may cause poorer performance in the *worst case* scenario due to adaptations to random fluctuations that can take time to reverse [2]. Potential negative implications may be mitigated by restricting the response-adaptive randomisation [2, 33]. Response-adaptive randomisation may be restricted in two ways. First, by imposing minimum and maximum allocation probabilities, which may be rescaled when arms are dropped. Second, by *softening*, i.e., raising the raw allocations probabilities to some exponent (the softening factor), which could be between 0 (leading to equal allocation probabilities after rescaling) and 1 (no restriction), most commonly between 0.5 and 1.0 [2, 50]. With a common control, a relatively higher and possibly fixed allocation probability to the control arm may increase statistical power [2, 36].

When response-adaptive randomisation is used, operational complexity is increased by the need to handle potential *time drift*. Time drift is the potential bias due to changes between periods in the included population or concurrent interventions used with different allocation probabilities [2, 10, 23, 33, 51]. Similarly, stratified block randomisation to balance important prognostic factors is difficult to combine with response-adaptive randomisation [2, 52]. For these reasons, using an adequate *burn-in* period before allowing response-adaptive randomisation *and* restricting the response-adaptive randomisation is advisable.

In *adaptr*, response-adaptive randomisation is based on each arm's overall probability of being best [2, 23, 53]. Softening factors can vary across adaptive analyses, to ensure, e.g., equal allocation or more restrictive response-adaptivity early in the trial. *adaptr* supports multiple specific control-arm allocation rules: ratios of 1 (for each non-control arm) to the square root to the number of non-control arms (for the control arm) [2, 36, 54] or a control arm allocation probability equal to the highest probability among the non-control arms [2, 18]. Fixed allocation probabilities may be used for some or all arms. Here, the example trial will initially use equal allocation probabilities of 33.3% to each arm, followed by response-adaptive randomisation with restrictions in the form of 25% minimum limits that will be rescaled when an arm is dropped, and a softening factor of 0.5:

```
setup_trial_binom(
  ...
  start_probs = c(1/3, 1/3, 1/3),
  fixed_probs = NULL,
  min_probs = c(0.25, 0.25, 0.25),
  rescale_probs = "limits",
  soften_power = 0.5,
  ...
)
```

*Analysis model and priors*
The statistical model, including priors, for the primary outcome in the *actual* trial should guide the selection of the statistical model used in simulations, although it is common and acceptable to simplify both the model and the estimation method. For example, adjusting for important covariates during simulations is complex and usually omitted, akin to conventional sample size calculations [2], and simulations may use conjugate models [27, 55] instead of full Markov chain Monte Carlo estimation, to ensure speed and feasibility.

*adaptr* supports different models and modelling approaches and only requires that draws from the posterior probability distributions are returned for each trial arm on the natural (absolute) scale for the outcome of interest [27]. Thus, for binary outcomes, posterior draws should reflect event probabilities in each arm. The number of posterior draws used in each arm should be adequate to compare trial arms; if, e.g., stopping thresholds are calibrated (section *"Calibration"*), a larger number may be required as it determines the granularity of the estimated probabilities (e.g., with 1,000 posterior draws, the minimum non-zero difference in probabilities is 0.1%-points). Thus, we use 10,000 posterior draws in the example. As we use `setup_trial_binom()` we do not manually have to specify an analysis model or priors; conjugate beta-binomial models with flat priors are used [35, 55]:

```
setup_trial_binom(
    ...
    n_draws = 10000,
    ...
)
```

*Stopping and arm dropping rules*
A maximum sample size must be specified for simulations and will affect multiple performance metrics, including the overall type 1 error rate and power [2]. Stopping and arm dropping rules for superiority, inferiority, practical equivalence, and futility may be specified in *adaptr* [2, 27], as illustrated in **Fig. 3**.

**Fig. 3**

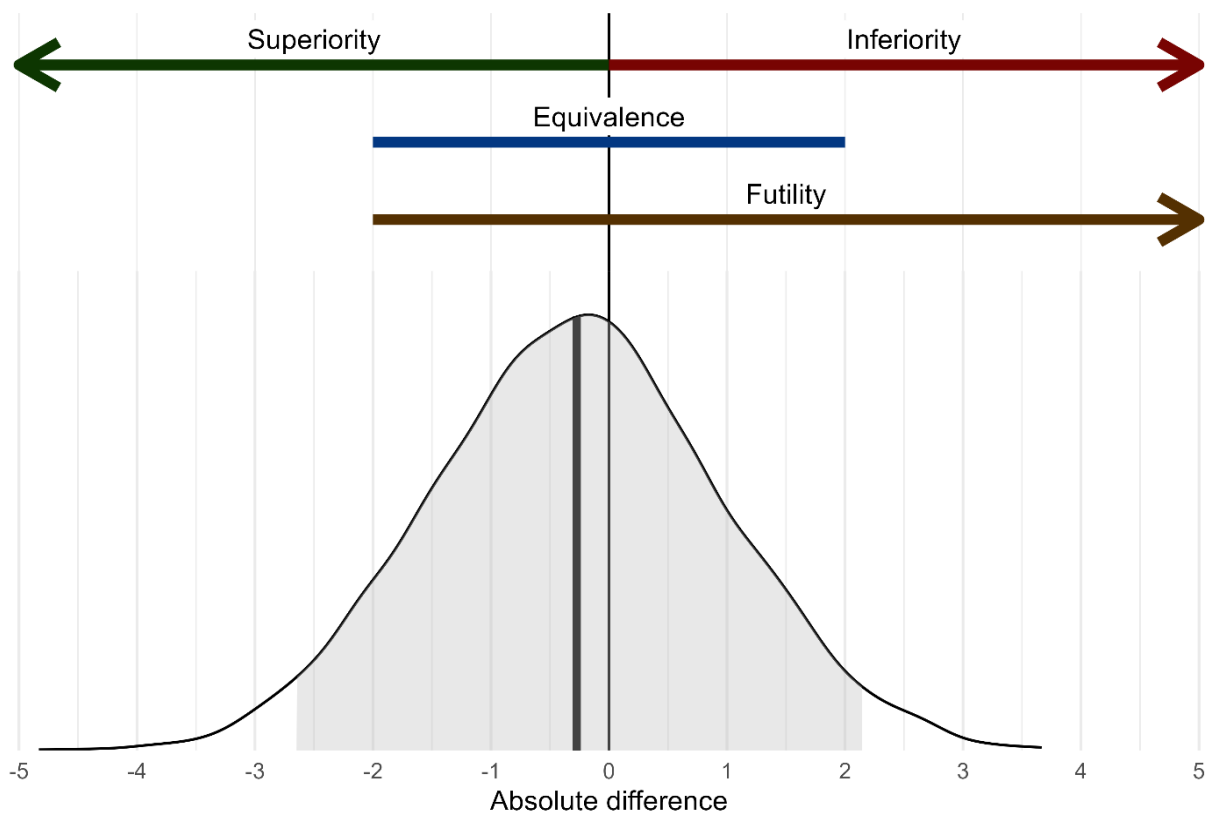

Figure legend: Illustration of probabilistic decisions rules for a single two-arm comparison with an undesirable outcome (i.e., negative differences are preferable). The lower part of the figure shows the posterior probability distribution on the absolute scale (e.g., %-points in the example used in the text) with the median value highlighted by the vertical bold line and the 95% percentile-based credible interval highlighted in grey. The upper part of the figure illustrates how the posterior is partitioned to calculate the probabilities of superiority, inferiority, practical equivalence, and futility, which are simply the proportion of posterior samples in each "*region*" of interest. Figure based on a similar figure previously presented elsewhere [2].

Stopping rules for superiority/inferiority are mandatory and have the highest priority, i.e., they will be assessed *before* stopping rules for practical equivalence or futility, as concluding that an arm is superior is more clinically useful than, e.g., a futility decision. Although stopping rules may be set so as to be sufficient to change clinical practice [2, 56], regulatory bodies will typically require type 1 error rates ≤5% [2, 25, 26, 57]. This can be achieved by manual iteration or automatic calibration (section *"Calibration"*). Stopping thresholds for superiority and inferiority are usually symmetric (i.e., the decision threshold for inferiority is defined as 100% minus the decision threshold for superiority) and may either be *constant* throughout the trial or more conservative at earlier analyses. The former correspond to Pocock and the latter to, e.g., O'Brien Fleming monitoring boundaries in conventional group sequential trial designs [33, 58].

Constant decision thresholds generally lead to smaller expected sample sizes, lower errors in estimates, and less overestimation of intervention effects when stopped early, but lower power compared to varying, decreasingly strict decision thresholds [59–61]. Consequently, the latter are often favoured in conventional trials *expected* to run until the maximum sample size and mainly use interim analyses as a safety measure, while constant decision thresholds may be preferable in advanced adaptive trials *not* expected to run until the maximum sample size.

Optional stopping rules for practical equivalence may be defined and will be evaluated *after* superiority/inferiority [2, 27]. Without a common control, the entire trial will be stopped if the largest absolute difference between all active arms is smaller than a pre-specified threshold with a sufficiently high probability, e.g., >90% probability that the largest absolute difference is <2.5%-points [2, 27]. With a common control, non-control arms will be dropped for equivalence if the absolute difference compared to the common control is smaller than a pre-specified threshold with sufficiently high probability; the overall trial will be stopped if only the common control remains.

Optional stopping rules for futility may be defined when a common control arm is used and will be evaluated *after* all other stopping rules. Non-control arms will then be dropped for futility if the probability that they are *not* sufficiently *better* than the control is above a pre-specified threshold, e.g., >90% probability of a *beneficial* difference is <2.5%-points, including the probability of the non-control arm being worse [2, 27]. The overall trial will be stopped if only the control remains.

In *adaptr*, all probability thresholds may vary across analyses and can be stricter in earlier analyses and more lenient in later analyses. By setting thresholds to either 100% or 0% (which will never be exceeded), stopping rules can be disabled at early analyses, making it possible to only use *some* stopping rules early or to use response-adaptive randomisation before allowing stopping or arm dropping. Of note, the *differences* of interest on the absolute scale for practical equivalence and futility stopping rules must be constant. Probability thresholds for practical equivalence and futility may be manually or automatically calibrated to obtain specific performance metrics, but this is optional. Probability thresholds for practical equivalence and futility may be lower than the corresponding superiority/inferiority thresholds as they will otherwise often require substantially more participants to be triggered [2]. Of note, allowing trials to stop for practical equivalence or futility may decrease the probability of stopping for superiority, and may thus reduce power [2]. With a common control, practical equivalence and/or futility may be evaluated only against the *first* control arm (as will often be most relevant), or also against any other arms that are subsequently promoted to controls.

Finally, if no maximum sample size is desired for the actual trial and an artificially high maximum sample size is specified (as described in section *"Analysis timing and outcome-data lag"*), the chosen stopping rules must lead to probabilities of triggering a stopping rule of 100% across all evaluated scenarios and sensitivity analyses of assumed parameters (section *"Sensitivity analyses of assumed parameters"*) to ensure valid estimates of trial design performance.

Here, we specify constant inferiority and superiority stopping thresholds and a stopping rule for equivalence that first becomes active when data from 1,500 participants are analysed. Following this, the trial stops if there is >90% probability that the largest absolute difference between all remaining arms is <2.5%-points:

```
setup_trial_binom(
  ...
  inferiority = 0.01,
  superiority = 0.99,
  equivalence_prob = ifelse(seq(from = 500, to = 10000, by = 250) < 1500, 1, 0.9),
  equivalence_diff = 0.025,
  ...
)
```

*Clinical scenarios*

Typically, advanced adaptive trial designs are evaluated under multiple clinical scenarios, e.g., assuming different arm-specific outcome distributions and thus different intervention effects [2]. Trial designs should be evaluated using a so-called *null* scenario *without* between-arm differences and at least one scenario *with* between-arm differences [2]. The probability of stopping for superiority in the *null* scenario corresponds to the type 1 error rate [2, 18, 24, 26, 62]. When desired, calibration will typically use the *null* scenario (see section *"Calibration"*). The primary *null* scenario should be based on the most likely *reference* outcome distribution, e.g., the most likely event probability based on existing clinical knowledge.

The probabilities of stopping for superiority in scenarios with differences are used to assess their *power* [2] and other performance metrics. We recommend using at least two scenarios with differences present, e.g., reflecting *small* and *large* differences, with at least one arm using the same outcome distribution as the primary *null* scenario. Small differences might be aligned with the thresholds used for equivalence or futility, and ideally correspond to the minimally relevant difference. Large differences might correspond to the anticipated or largest expected realistic intervention effect [2].

Practically, each scenario is expressed in *adaptr* as one trial design specification with different outcomes in each arm and all other design choices being identical across scenarios, which are then evaluated. The full code to specify the primary *null scenario* is shown here and combines the code snippets presented so far with fewer comments:

```r
primary_design_null_scenario <- setup_trial_binom(
  # Arms and scenario
  arms = c("Arm A", "Arm B", "Arm C"),
  control = NULL,
  true_ys = c(0.25, 0.25, 0.25),
  highest_is_best = FALSE,
  # Allocation rules
  start_probs = c(1/3, 1/3, 1/3),
  fixed_probs = NULL,
  min_probs = c(0.25, 0.25, 0.25),
  rescale_probs = "limits",
  soften_power = 0.5,
  # Participants with data available/randomised at each analysis
  data_looks = seq(from = 500, to = 10000, by = 250),
  randomised_at_looks = c(seq(from = 700, to = 9950, by = 250), 10000),
  # Stopping rules
  inferiority = 0.01,
  superiority = 0.99,
  equivalence_prob = ifelse(seq(from = 500, to = 10000, by = 250) < 1500, 1, 0.9),
  equivalence_diff = 0.025,
  # Posterior draws
  n_draws = 10000
)
```

**Performance metrics**

Performance metrics of interest must be chosen and prioritised before simulating and comparing design variants. Different metrics (**Table 1**) may be preferred according to the research question and specific trial [2, 18, 24], and optimising one performance metric will often worsen other metrics.

**Table 1 (continued on next page)**

| Performance metric | Description |
|---|---|
| Sample size | Total sample size (across arms) in each simulation. Summarised across simulations using means (i.e., expected values), SDs, medians, IQRs, and ranges. Lower sample sizes are preferable for economical/logistical reasons and to allow results to be used faster for future patients. Note that a low mean sample size does not rule out a small probability of a very large sample size; one should therefore look at all the sample size metrics. |
| Summed outcome data | Total summed outcome data (across all arms) in each simulation, i.e., total event counts for binary outcomes (e.g., mortality) or total sums of continuous outcomes (e.g., days alive and out of hospital). Summarised across simulations using means, SDs, medians, IQRs, and ranges. Depending on whether an undesirable or desirable outcome is used, lower or higher values, respectively, are preferable for *internal* patients (trial participants). |
| Ratio of summed outcome data to sample size | Ratio of total summed outcome data to sample size (across arms; summed outcome data divided by sample size) in each simulation, i.e., total event probabilities for binary outcomes (e.g., mortality) or grand means for continuous outcomes (e.g., days alive and out of hospital). Summarised across simulations using means, SDs, medians, IQRs, and ranges. Depending on whether an undesirable or desirable outcome is used, lower or higher values, respectively, are preferable for *internal* patients (trial participants). |
| Probabilities of conclusiveness, superiority, equivalence, futility, and stopping after the maximum number of adaptive analyses without triggering any stopping rule | The proportions of simulated trials stopped due to different stopping rules, i.e., superiority, practical equivalence, and futility. The probability of conclusiveness is the combined probability of stopping for either superiority, practical equivalence, or futility, while the probability of stopping after the maximum number of adaptive analyses without triggering any stopping rule corresponds to the probability of inconclusiveness. The probability of superiority may be interpreted as the Bayesian analogue to the type 1 error rate for scenarios containing no between-arm outcome differences, and as the Bayesian analogue of the power for scenarios containing between-arm differences [2]. It is typically necessary to keep the type 1 error rate at or below a specific value (e.g., 5%) [25, 26] and stopping rules may be *calibrated* to this target value; higher probabilities of conclusiveness are preferred (especially with regards to *external*/future patients) to increase the usefulness of trial results, and higher probabilities of superiority may be desired at the expense of lower probabilities of equivalence or futility, as superiority decisions may be more clinically useful if a difference exists. |
| Probabilities of selecting different arms or no arm | The proportions of simulated trials selecting different arms or no arm after stopping (according to the arm selection strategy used [2], as described in the text). For both *internal* patients (participants) and *external*/future patients, higher probabilities of selecting arm(s) that are truly better are preferable. |
| Root mean squared error/median absolute error of the effect estimate in the selected arm | Root mean squared error (RMSE)/median absolute error (MAE) of the intervention effect estimate (e.g., the estimated event probability) for the selected arm in each simulation (according to the arm selection strategy used, as described in the text [2]) across simulations compared to the 'true' simulated value across trials. Lower RMSEs/MAEs are preferable as this corresponds to higher accuracy. |

| Performance metric | Description |
|---|---|
| Root mean squared error/median absolute error of the intervention effect | Root mean squared error (RMSE)/median absolute error (MAE) of the estimate of the intervention effect (the difference between the estimate for a selected non-control arm compared to the estimate from a control or otherwise specified reference arm) compared to the 'true' intervention effect (the difference between the 'true' simulated value for the selected non-control arm compared to the 'true' simulated value in a control/reference arm) across trials. Lower RMSEs/MAEs are preferable as this corresponds to higher accuracy. Calculation depends on the arm selection strategy used, as described in the text [2]. |
| Ideal design percentage | The ideal design percentage (IDP) [2, 18] combines arm selection probabilities and the consequences of selecting different arms into a single measure (e.g., high probabilities of selecting an arm that is only slightly inferior to the best arm is less problematic than high probabilities of selecting an arm that is substantially worse than the best arm). Especially relevant when comparing trial designs with >2 arms; higher values are preferable as this corresponds to increased benefit for *external*/future patients. Calculation depends on the arm selection strategy used as described in the text [2]; IDPs are not calculated for scenarios with no between-arm differences. Mostly relevant and interpretable when comparing multiple designs. |

Table legend: Performance metrics automatically calculated by *adaptr*, adapted from [2]. Other performance metrics may also be of interest in specific cases, but these are not automatically calculated by *adaptr* and not covered here. Abbreviations: IDP: ideal design percentage; IQR: interquartile range (i.e., 25% and 75% percentiles); MAE: median absolute error; RMSE: root mean squared error; SD: standard deviation.

Some performance metrics (arm selection probabilities, error metrics, and ideal design percentages) are calculated based on the *selected* arms. Different *arm selection strategies* may be chosen for simulations where superiority is not concluded, based on, e.g., which arm would be used in clinical practice afterwards if the trial is inconclusive [2]. For example, these performance metrics may be calculated in the following ways:

- For trial simulations ending with superiority only.
- For simulations not ending with superiority:
    - Considering the original common control arm (if any and if not dropped previously) selected in simulations not ending with superiority.
    - Considering a specific arm (i.e., the cheapest/most available intervention) selected in simulations not ending with superiority.
    - Considering the arm with the highest probability of being best at the last analysis selected in simulations not ending for superiority.

Some metrics may be prioritised for logistical/economic reasons (e.g., mean sample sizes), to maximise benefits to *participants* (e.g., total event counts and event probabilities), to maximise benefits to *external* and future patients (probabilities of conclusiveness/superiority, arm selection probabilities, ideal design percentages [2, 18]) and to maximise accuracy of the trial results (error metrics). The probabilities of superiority and conclusiveness and the expected sample sizes will typically be of high priority.

**Simulations and performance metric calculation**
Many trial simulations are needed to accurately assess performance, and, possibly, compare multiple candidate designs. Performance may be assessed directly, followed by manually revising and re-assessing trial designs in an iterative process until performance metrics are acceptable. Alternatively, an automatic process may be used to *calibrate* a specific design parameter to obtain an acceptable value for a specific performance metric (section *"Calibration"*). Typically, ensuring that stopping rules for superiority/inferiority lead to acceptable type 1 error rates is central. Manual assessment may be carried out by conducting simulations under the *primary null* scenario with specified stopping rules and assessing whether the overall type 1 error and other metrics are acceptable. If the type 1 error rate is too high, stopping thresholds may be made more restrictive followed by a new round of simulations in an iterative fashion until the overall type 1 error is acceptable. Similarly, if the type 1 error rate is below the desired value, stopping rules may be made more lenient followed by repeated simulations and evaluation, to decrease the expected sample size. Alternatively, this may be done using an automatic calibration procedure (section *"Calibration"*). Other elements of the trial design may also be iteratively revised to ensure that other performance metrics are acceptable. Conducting 100,000 simulations is generally recommended when evaluating type 1 error rates [26], but fewer (e.g., 10,000) simulations may be enough for evaluating other metrics where less precision is required, or if uncertainty measures (e.g., 95% confidence intervals [CIs]) are calculated and found acceptable [26].

In *adaptr,* trial simulations may be conducted using `run_trials()`. When simulations have been conducted, results may be calculated, extracted, and summarised using multiple functions. `extract_results()` returns data in a tabular (`data.frame`) format with 1 row per simulation and 1 column per data point. `check_performance()` summarises performance metrics across trials in a `data.frame` format with optional calculation of uncertainty measures, e.g., 95% CIs, calculated using non-parametric bootstrapping with resampling with replacement of the results obtained from the individual simulations [63]. Finally, `summary()` calculates performance metrics and summarises simulation results in a *list* format with a dedicated print method.

In the example, we conduct 10,000 simulations using the previously defined trial specification under the *primary null* scenario (section *"Clinical scenarios"*). This is followed by calculation of performance metrics with uncertainty measures in the form of 95% CIs while considering the "best" remaining arm (i.e., the highest probability of being superior as described in section *"Performance metrics"*) selected in simulations not stopped for superiority:

```
primary_sims_uncalibrated <- run_trials(
  trial_spec = primary_design_null_scenario,
  n_rep = 10000,
  base_seed = 4131, # Reproducibility
  path = paste0(dir_out, "Primary sims uncalibrated.RDS") # Save/reload
)

primary_performance_uncalibrated <- check_performance(
  primary_sims_uncalibrated,
  select_strategy = "best",
  uncertainty = TRUE,
  n_boot = 5000, # Number of bootstrap resamples
  ci_width = 0.95, # 95% CIs
  boot_seed = 4131 # Reproducibility
)
```

All performance metrics are included in **appendix 1**; the percentage of simulations stopped for superiority, i.e., the type 1 error rate in this case, is 5.3% (95% CI: 4.8% to 5.7%), with a mean sample size of 7,881 participants. Even though less simulations are conducted than the 100,000 recommended, the actual estimate and the 95% CI of indicates that the type 1 error rate is likely to exceed the typically recommended 5%. Consequently, to ensure an acceptable type 1 error rate, it may be necessary to change the stopping rules or at least conduct a larger number of simulations to decrease the uncertainty, in which case the type 1 error rate could turn out to be acceptable.

**Calibration**
To adequately control the overall type 1 error rate for the guiding outcome, stopping rules for superiority/inferiority may be automatically calibrated. Calibration according to a specific desired type 1 error rate should be performed under the *primary null scenario* and combines repeated simulations with an algorithm that aims to find stopping thresholds that achieves the desired type 1 error rate.

*adaptr* supports automatic calibration of trial specifications to obtain constant, symmetrical stopping rules for superiority/inferiority that target the typically recommended type 1 error rate of 5% [2, 25, 26, 57], but also supports calibration of non-constant/non-symmetric stopping rules or other design choices to optimise another performance metric. *adaptr* uses a Gaussian process-based Bayesian optimisation algorithm [64] that aims to efficiently (i.e., with as few sets of simulations as possible) identify stopping rules that will lead to the desired type 1 error rate. Below, we calibrate the superiority and inferiority stopping rules using `calibrate_trial()` with 10,000 simulations in each calibration step. A *target* value for the type 1 error rate (5%), a tolerance threshold and direction of tolerance, a search range for the stopping threshold for superiority (with the threshold for inferiority defined as 1 - the threshold for superiority in this case), along with a maximum number of iterations allowed determines when the calibration procedure is stopped:

```
primary_design_null_scenario_calibration <- calibrate_trial(
  trial_spec = primary_design_null_scenario,
  n_rep = 10000,
  base_seed = 4131, # Reproducibility
  # Target, search range, tolerance, and maximum number of iterations
  target = 0.05,
  search_range = c(0.9, 1),
  tol = 0.001,
  dir = -1, # Only tolerate values below target, i.e., 0.049 to 0.050
  iter_max = 25,
  path = paste0(dir_out, "Primary calibration.RDS") # Save/reload
)
```

*Evaluating the calibration procedure and results*
Following the calibration process, it should be checked if the calibration procedure was successful, i.e., that an acceptable type 1 error rate was achieved within the maximum permitted number of iterations. If not, consider using more posterior draws, more iterations, a wider search range, or a wider tolerance range for the target value. Example code to summarise calibration results and extract simulations and other data following calibration, including the relevant outputs, is included in **appendix 1**.
In the example, the calibration was successful, with a resulting stopping threshold for superiority of 0.990416 (rounded to six significant digits and corresponding to a stopping threshold for inferiority of 0.009584). As only 10,000 simulations were conducted during each iteration in the calibration process, uncertainty measures should be calculated and checked, *or* alternatively the final calibrated trial design should be evaluated using 100,000 simulations to ensure that the type 1 error rate (and other performance metrics) remain acceptable. For practical purposes, stopping rules with a limited number of digits are easier to use, however, rounding requires new evaluation. Below, the calibrated stopping rules are rounded to four significant digits, followed by conduct of 100,000 simulations and performance evaluation:

```r
# Extract and round calibrated stopping rule for superiority ('best_x')
superiority_rounded <- round(primary_design_null_scenario_calibration$best_x, 4)

# Extract calibrated trial design specification ('best_trial_spec') and update
# to use rounded stopping rules (inferiority = 1 - superiority)
primary_design_null_scenario_calib <-
  primary_design_null_scenario_calibration$best_trial_spec
primary_design_null_scenario_calib$superiority <- superiority_rounded
primary_design_null_scenario_calib$inferiority <- 1 - superiority_rounded

# Run large number of simulations with updated trial design specification
primary_null_calibrated <- run_trials(
  primary_design_null_scenario_calib,
  n_rep = 100000,
  path = paste0(dir_out, "Primary sims calibrated.RDS"), # Save/reload
  base_seed = 4131 # Reproducibility
)

# Check performance metrics without calculating uncertainty measures (not
# necessary due to the large number of simulations)
primary_performance_calibrated_rounded <- check_performance(
  primary_null_calibrated,
  select_strategy = "best"
)
```

All outputs are included in **appendix 1**; the probability of superiority (type 1 error rate) for these 100,000 simulations is 4.8%, with a mean sample size of 7,932 participants. As the type 1 error rate is still acceptable, we proceed with these stopping rules. Notably, the probability of conclusiveness is only 66.4% due to only 61.6% probability of stopping for practical equivalence in addition to the 4.8% probability of stopping for superiority. This may be too low, and increasing the maximum sample size or making the stopping rule for practical equivalence more lenient should be considered in cases like this before proceeding with evaluations under other clinical scenarios.

**Performance metric assessment under other clinical scenarios**
Following successful calibration and acceptable results under the *null scenario*, the trial design and calibrated stopping rules may be used to conduct additional simulations evaluating the design under other scenarios as described in section *"Clinical scenarios"*. **Table 2** contains selected performance metrics for the example trial design evaluated under 15 example scenarios constituting the unique combinations of *small* differences, corresponding to the threshold for practical equivalence, and *large* differences, corresponding to two times the equivalence threshold, in both directions (all performance metrics and the corresponding code is included in **appendix 1**).

**Table 2**

| Arm A | Arm B | Arm C | Mean sample size | Probability of conclusiveness[1] | Probability of superiority[2] | Probability of equivalence[3] |
|-------|-------|-------|------------------|----------------------------------|-------------------------------|-------------------------------|
| 25.0% | 25.0% | 25.0% | 7932 | 66.4% | 4.8% | 61.6% |
| 25.0% | 27.5% | 25.0% | 6496 | 85.2% | 14.6% | 70.6% |
| 25.0% | 22.5% | 25.0% | 6473 | 81.5% | 59.7% | 21.8% |
| 25.0% | 30.0% | 25.0% | 5304 | 97.1% | 14.1% | 83.0% |
| 25.0% | 20.0% | 25.0% | 2871 | 100.0% | 99.6% | 0.4% |
| 25.0% | 27.5% | 27.5% | 6710 | 77.5% | 56.5% | 21.1% |
| 25.0% | 22.5% | 27.5% | 5052 | 95.2% | 74.3% | 20.9% |
| 25.0% | 30.0% | 27.5% | 5287 | 93.1% | 73.0% | 20.1% |
| 25.0% | 20.0% | 27.5% | 2350 | 100.0% | 99.8% | 0.2% |
| 25.0% | 22.5% | 22.5% | 6239 | 87.2% | 13.7% | 73.6% |
| 25.0% | 30.0% | 22.5% | 4716 | 96.4% | 74.6% | 21.8% |
| 25.0% | 20.0% | 22.5% | 4788 | 96.7% | 75.5% | 21.2% |
| 25.0% | 30.0% | 30.0% | 3205 | 99.8% | 99.3% | 0.5% |
| 25.0% | 20.0% | 30.0% | 2211 | 100.0% | 99.8% | 0.2% |
| 25.0% | 20.0% | 20.0% | 4650 | 99.0% | 13.8% | 85.1% |

Table legend: Event probabilities in each arm and selected performance metrics under the 15 clinical scenarios evaluated. The scenarios have constant assumed event rates of 25.0% in one arm and varying event probabilities and combinations in the other arms, corresponding to the primary *null* scenario and the unique combinations of no/small/large differences in both directions. We conducted 100,000 simulations of the scenario without differences between arms but only 10,000 simulations for each scenario with between-arm differences, as these are not used for calibrating the stopping rules and as less accuracy for the other performance metrics will often be acceptable than for type 1 error rates. Uncertainty measures for all metrics can be calculated if required and is mostly relevant with <100,000 simulations (see section *"Simulations and performance metric calculation"*).

[1] Probability of triggering any stopping rule at or before the maximum allowed sample size, i.e., the probability of either superiority or practical equivalence in this case.

[2] The probability of superiority corresponds to the type 1 error rate in the scenario without differences between arms and the power in all scenarios with differences between arms [2].

[3] The probability of equivalence refers to the final decision between all arms remaining at the last conducted analysis. In this example, one arm may be dropped for inferiority early and the remaining two arms may then be declared practically equivalent at a later analysis. Proportions of various combinations of available arms in the last analysis conducted in a set of simulations can be summarised using the `check_remaining_arms()` function in *adaptr* [27].

The probabilities of conclusiveness across the 15 scenarios ranged from 66.4% to approximately 100%, with mean sample sizes ranging from 2,211 to 7,932.

**Sensitivity analyses assessing of design variants**

Following the conduct of simulations for the initial trial design under multiple scenarios, additional design variants or rounds of iteration may be necessary to refine the design or compare the effect of different design choices. These choices include, e.g., the number and timing of analyses, the stopping rules (especially those not calibrated, i.e., the practical equivalence stopping rule in this case), the randomisation scheme (i.e., the use of fixed- or response-adaptive randomisation or combinations, and any restrictions used with response-adaptive randomisation), and the follow-up duration and data outcome-lag period for the guiding outcome. In the example (**Table 2**), it may be necessary to increase the maximum sample size or revise the design (including the stopping rules) to increase the probabilities of conclusive results across all clinical scenarios evaluated. Even when the results from the first set of simulations are considered acceptable, sensitivity analyses varying different key design parameters are recommended to assess their potential influence and as this could potentially further improve performance. To limit the number of required simulations compared to doing simulations for all combinations of different values for the key parameters, assessing design variants will typically be done by varying key parameters one at a time, possibly deciding on using a different design variant as the reference, and then eventually further varying other key parameters one at a time. Importantly, whenever *fixed* design parameters (i.e., those controlled by the trialists) are changed and assessed, the calibration process (if used) should generally be repeated; during this process, we recommend that all *assumed* but essentially uncontrollable parameters (e.g., inclusion rates and outcome distributions) are unchanged followed by assessment later (section *"Sensitivity analyses of assumed parameters"*). Suggested sensitivity analyses are outlined in **Table 3**.

**Table 3**

| Suggested sensitivity analyses of fixed design parameters |
|---|
| - Other stopping rules for superiority/inferiority (if relevant; mostly relevant if not calibrated) |
| - Rounding of stopping rules for superiority/inferiority (if calibrated) |
| - Other randomisation schemes (e.g., fixed randomisation, more/less restricted response-adaptive randomisation, different types or degrees of restriction [limits and/or softening factors]) |
| - Different comparison strategy if relevant (e.g., all-versus-all comparison if primary design has a common control arm or vice versa if a relevant common control arm may be specified; only relevant for designs with >2 arms) |
| - Different analysis timings (including burn-in) and/or maximum sample sizes |
| - Different stopping rules for practical equivalence or futility if used (different probability thresholds and stricter thresholds most relevant, but different thresholds over time may also be relevant; if relevant, different absolute thresholds for practical equivalence/futility) |
| - Different outcome-data lag periods (if relevant; i.e., different follow-up duration and/or different permitted lag period for data collection, cleaning, and verification) |
| **Suggested sensitivity analyses of assumed parameters** |
| - Different reference distributions (differences in both directions, e.g., different reference event probabilities; either a grid of different values or a range of values representing the range of *a priori* plausible values) |
| - Different inclusion rates (in both directions, either in a grid or a range of values representing what is *a priori* considered plausible; possibly different inclusion rates over time, e.g., if constant inclusion rates are assumed, challenging this may be considered) |

Table legend: Suggested sensitivity analyses for consideration when evaluating an advanced adaptive trial using simulations. For the sensitivity analysis of fixed design parameters, any calibration procedure should be repeated when these are changed (except for the evaluation of rounding calibrated stopping rules). For the sensitivity analyses of assumed parameters, the stopping rules should be identical to the primary evaluation, i.e., any calibration procedure should not be repeated here. Sensitivity analyses should be conducted using a set of scenarios corresponding to those used for the primary analyses.

**Sensitivity analyses of assumed parameters**
For at least the final design, sensitivity analyses should be conducted assessing the potential impact of the *assumed* parameters, while the *fixed* design parameters are kept constant (see **Table 3** for suggestions). It is especially important to assess the influence of different assumed reference outcome distributions, as this may affect all performance metrics, including type 1 error rates, power, and expected sample sizes [26]. Importantly, key performance metrics should be acceptable across a range or grid of values covering the plausible, different reference distributions [26]. We also recommend sensitivity analyses varying the assumed inclusion rates [37]. Sensitivity analyses covering the range of plausible values for each parameter should be conducted, and the resulting performance metrics should be acceptable under the full range of reasonably plausible assumptions. It is central that such sensitivity analyses are conducted using the *same* stopping rules as in the corresponding simulations conducted under the primary assumptions (as the stopping rules needs to be determined prior to trial start), i.e., if the stopping rules were calibrated under the primary *null scenario*, the exact same stopping rules resulting from that calibration should be used in the sensitivity analyses without recalibration.

**Reporting**

All design characteristics, assumptions, and performance metrics from both the primary simulations and sensitivity analyses of the final trial design should be reported as part of the trial protocol. Further, presenting results from sensitivity analyses of design variants and assumed parameters and earlier iterations of the trial design, even if not used, is recommended as this makes the decision of the final trial design transparent. An example from an actual trial protocol is available elsewhere [49]. Including the simulation code (including random seeds and software version info) when reporting results may be considered to increase transparency and allow replication and serve as an aide for other trialists planning similar designs [65, 66].

**Additional examples**

Additional examples, including of more customised designs using `setup_trial()`, are included in **appendix 2**:

- **Example 1** illustrates how to use a common control arm.
- **Example 2** illustrates how to use a more complex outcome distribution.
- **Examples 2 and 3** illustrate how to define a custom function to return posterior draws, which may use any estimation method and any desired priors.

Further examples, including examples explicitly considering inclusion rates and outcome-data lag in greater detail, are available elsewhere [37, 49].

# Discussion

We have provided a thorough example-based guide on the steps required for evaluating and comparing advanced adaptive Bayesian trial designs using adaptive stopping, arm dropping, and response-adaptive randomisation along with full simulation code using a well-documented and flexible simulation engine [27].

The primary strength of this manuscript is that we have covered the key methodological decisions needed when planning advanced adaptive trials from a theoretical and practical point of view, including providing the complete, annotated code covering the entire workflow. Given that limited guidance on this topic exists, we hope this will serve as a valuable reference for trialists considering or using these designs. The *adaptr* package used [27] has the benefits of being open-source, freely accessible, relatively easy to use, well-documented, extensible, and optimised to be relatively fast. However, other software packages for adaptive trial simulation exist [28, 33] and may also be considered.

This guide and our simulation engine come with some limitations. First, while we have aimed to provide comprehensive guidance on assessing and comparing advanced adaptive trials, not every adaptive feature is covered here or supported by the *adaptr* [27] package. Primarily, adaptive enrichment (restricting allocation to those most probable to benefit) [67] including separate adaptations in different subgroups [10], and adaptive arm adding (including *staggered entry*) as used in some adaptive platform trials [10] is not covered or supported by the package. However, this framework supports the planning of platform trials with interventions nested in *domains* (groups of comparable interventions similar to what could be compared in a stand-alone trial) as long as the platform only allows addition of new domains (which may include interventions assessed in previous domains), but not new interventions within domains (i.e., domains are *closed*) [1, 2, 10, 14]. Second, while we have provided guidance on assessing and comparing different advanced adaptive trial designs, we provide limited guidance on the choice of specific adaptive features beyond that they should be compared in each case using simulation. This is intentional, as some adaptive features may be beneficial in some situations while having undesirable effects in other situations; this balance also depends on the prioritisation of different performance metrics (e.g., response-adaptive randomisation may lower expected sample sizes in some trial designs and increase it in others [2], while increasing individual participants' chances of better outcomes in both scenarios). While some general intuition is provided, it is important not to blindly rely on this or previous results but evaluate key design choices specifically in each case. Further, it is not possible to provide universal guidance on the prioritisation of different performance metrics, as the prioritisation and acceptable trade-offs needs to be considered separately in each trial being planned. Third, limiting the scope somewhat was necessary, and we consequently primarily focused on late-phase, large and pragmatic Bayesian adaptive trials. As such, adaptive trials using frequentist statistical methods or adaptive trial designs used for earlier phase trials (including dose-finding trials) are not covered here. However, most of the considerations discussed also apply to such trial designs, even if the planning, evaluating, and final interpretation may differ somewhat. Fourth, the main text has only covered a design with no common control arm and using a binary outcome with default, flat priors; however, the *adaptr* package documentation and **appendix 2** provides examples on how to specify simulations using other outcome types, custom priors, common control arms, and other of the more complex features of *adaptr* [27] and the rest of the workflow is similar.

# Conclusions

In conclusion, this practical guide provides comprehensive advice for trialists considering or planning advanced adaptive Bayesian trials using adaptive stopping, arm dropping, and response-adaptive randomisation. By including examples of simulation-based trial design assessment and comparison, we have covered not only the methodological considerations, but also the practical aspects of doing simulation-based trial design evaluation and comparison. While planning and simulating advanced adaptive trials is an iterative process that typically will be more time-consuming than designing and planning conventional trials, the additional effort in the planning phase will often be outweighed by higher flexibility, increased effectiveness, and higher probabilities of conclusiveness in the resulting trials.

# Declarations

**Availability of data and materials**
This study is based on simulated data only. The complete, annotated analysis code used to generate the simulated data is available in the supplementary information.

**Conflicts of interest**
None.

**Funding**
This study was conducted as part of the *Intensive Care Platform Trial* (*INCEPT*) research programme (www.incept.dk), which is funded by Sygeforsikringen 'danmark' and the Novo Nordisk Foundation, with additional support from Grosserer Jakob Ehrenreich og Hustru Grete Ehrenreichs Fond, Dagmar Marshalls Fond, and Savværksejer Jeppe Juhl og hustru Ovita Juhls Mindelegat and domain-specific funding from Danmarks Frie Forskningsfond. None of the funders had any influence on any aspects of this study.

**Author's contributions**
Conceptualization: AG, AKGJ, BSKH. Data curation: AG. Formal analysis: AG. Funding acquisition: AG, TL, AP, MHM. Investigation: AG. Methodology: all authors. Project administration: AG. Software: AG, AKGJ, TL, BSKH. Visualization: AG. Writing – original draft: AG. Writing – review and editing: all authors.

# Supplementary information

The complete, annotated simulation code used for the primary example along with all outputs are included in **appendix 1**. Description and code for additional examples are included in **appendix 2**.

# Appendix 1

This supplementary appendix includes the complete simulation code used for the primary example in the manuscript along with the outputs.

For additional information on using the ***adaptr*** package and on the package outputs, please see the complete package documentation available at: https://inceptdk.github.io/adaptr/.

**Setup**

Below, the package is loaded and a cluster for parallel computation initiated (the number of cores used may be changed by the user as needed and considering the number of cores available). In addition, the directory used to save simulations and results in is specified (**note: an actual path must be inserted here**). This is done to avoid having to re-run simulations when no changes have been made.

```r
library(adaptr)

## Loading 'adaptr' package v1.4.0.
## For instructions, type 'help("adaptr")'
## or see https://inceptdk.github.io/adaptr/.

setup_cluster(10) # Number of cores for parallel computation
dir_out <- "<PATH>/" # Replace with an actual, permanent path
```

**Code snippets**

The following section includes all code snippets included in the main text of the manuscript as part of the setup of the trial specification (under subheadings corresponding to those in the main text).

These snippets are incomplete and hence not run before the full trial specification is provided below.

*Interventions and use of common control*

```r
setup_trial_binom(
  arms = c("Arm A", "Arm B", "Arm C"),
  control = NULL,
  ...
)
```

*Outcome type and generation*

```r
setup_trial_binom(
  ...
  true_ys = c(0.25, 0.25, 0.25),
  highest_is_best = FALSE,
  ...
)
```

*Analysis timing and outcome-data lag*

```
setup_trial_binom(
  ...
  # Number of participants with data available and included in each analysis
  data_looks = seq(from = 500, to = 10000, by = 250),
  # Number of participants randomised at each analysis
  randomised_at_looks = c(seq(from = 700, to = 9950, by = 250), 10000),
  # Note: the maximum number in both arguments should be equal
  ...
)
```

*Allocation profiles*

```
setup_trial_binom(
  ...
  start_probs = c(1/3, 1/3, 1/3),
  fixed_probs = NULL,
  min_probs = c(0.25, 0.25, 0.25),
  rescale_probs = "limits",
  soften_power = 0.5,
  ...
)
```

*Analysis model and priors*

```
setup_trial_binom(
  ...
  n_draws = 10000,
  ...
)
```

*Stopping and arm dropping rules*

```
setup_trial_binom(
  ...
  inferiority = 0.01,
  superiority = 0.99,
  equivalence_prob = ifelse(seq(from = 500, to = 10000, by = 250) < 1500, 1, 0.9),
  equivalence_diff = 0.025,
  ...
)
```

**Complete trial design specification**

Here, all the incomplete snippets are combined into a complete trial specification (with fewer comments), which can be run:

```r
# Specification
primary_design_null_scenario <- setup_trial_binom(
  # Arms and scenario
  arms = c("Arm A", "Arm B", "Arm C"),
  control = NULL,
  true_ys = c(0.25, 0.25, 0.25),
  highest_is_best = FALSE,
  # Allocation rules
  start_probs = c(1/3, 1/3, 1/3),
  fixed_probs = NULL,
  min_probs = c(0.25, 0.25, 0.25),
  rescale_probs = "limits",
  soften_power = 0.5,
  # Participants with data available/randomised at each analysis
  data_looks = seq(from = 500, to = 10000, by = 250),
  randomised_at_looks = c(seq(from = 700, to = 9950, by = 250), 10000),
  # Stopping rules
  inferiority = 0.01,
  superiority = 0.99,
  equivalence_prob = ifelse(seq(from = 500, to = 10000, by = 250) < 1500, 1, 0.9),
  equivalence_diff = 0.025,
  # Posterior draws
  n_draws = 10000
)

# Print design specification
primary_design_null_scenario

## Trial specification: generic binomially distributed outcome trial
## * Undesirable outcome
## * No common control arm
## * Best arms: Arm A and Arm B and Arm C
##
## Arms, true outcomes, starting allocation probabilities
## and allocation probability limits (min/max_probs rescaled):
##    arms true_ys start_probs fixed_probs min_probs max_probs
## Arm A    0.25       0.333          NA       0.25        NA
## Arm B    0.25       0.333          NA       0.25        NA
## Arm C    0.25       0.333          NA       0.25        NA
##
## Maximum sample size: 10000
## Maximum number of data looks: 39
## Planned data looks after:  500, 750, 1000, 1250, 1500, 1750, 2000, 2250, 2500, 2750,
3000, 3250, 3500, 3750, 4000, 4250, 4500, 4750, 5000, 5250, 5500, 5750, 6000, 6250, 650
0, 6750, 7000, 7250, 7500, 7750, 8000, 8250, 8500, 8750, 9000, 9250, 9500, 9750, 10000
patients have reached follow-up
## Number of patients randomised at each look:  700, 950, 1200, 1450, 1700, 1950, 2200,
2450, 2700, 2950, 3200, 3450, 3700, 3950, 4200, 4450, 4700, 4950, 5200, 5450, 5700, 595
0, 6200, 6450, 6700, 6950, 7200, 7450, 7700, 7950, 8200, 8450, 8700, 8950, 9200, 9450,
9700, 9950, 10000
##
## Superiority threshold: 0.99 (all analyses)
```

```
## Inferiority threshold: 0.01 (all analyses)
## Equivalence thresholds:
## 1, 1, 1, 1, 0.9, 0.9, 0.9, 0.9, 0.9, 0.9, 0.9, 0.9, 0.9, 0.9, 0.9, 0.9, 0.
## 9, 0.9, 0.9, 0.9, 0.9, 0.9, 0.9, 0.9, 0.9, 0.9, 0.9, 0.9, 0.9, 0.9, 0.9,
## 0.9, 0.9, 0.9
## (no common control)
## Absolute equivalence difference: 0.025
## No futility threshold (not relevant - no common control)
## Soften power for all analyses: 0.5
```

**Simulations and performance metric evaluation without calibration**

```r
primary_sims_uncalibrated <- run_trials(
  trial_spec = primary_design_null_scenario,
  n_rep = 10000,
  base_seed = 4131, # Reproducibility
  path = paste0(dir_out, "Primary sims uncalibrated.RDS") # Save/reload
)

primary_performance_uncalibrated <- check_performance(
  primary_sims_uncalibrated,
  select_strategy = "best",
  uncertainty = TRUE,
  n_boot = 5000, # Number of bootstrap resamples
  ci_width = 0.95, # 95% CIs
  boot_seed = 4131 # Reproducibility
)

# Print and save
primary_performance_uncalibrated
```

```
##                metric        est  err_sd  err_mad        lo_ci        hi_ci
## 1         n_summarised  10000.000   0.000    0.000   10000.000   10000.000
## 2            size_mean   7880.700  24.402   23.900    7831.789    7928.992
## 3              size_sd   2436.432  14.790   14.670    2407.666    2465.283
## 4          size_median   8950.000   6.122    0.000    8950.000    8950.000
## 5            size_p25   5450.000 119.514    0.000    5450.000    5700.000
## 6            size_p75  10000.000   0.000    0.000   10000.000   10000.000
## 7             size_p0    700.000      NA       NA         NA         NA
## 8           size_p100  10000.000      NA       NA         NA         NA
## 9         sum_ys_mean   1969.661   6.105    5.992    1957.485    1981.875
## 10          sum_ys_sd    610.416   3.708    3.695     603.144     617.657
## 11      sum_ys_median   2230.000   8.779    8.896    2215.500    2249.000
## 12         sum_ys_p25   1394.000  14.147   11.305    1367.000    1425.000
## 13         sum_ys_p75   2484.000   1.169    1.483    2482.000    2486.000
## 14          sum_ys_p0    157.000      NA       NA         NA         NA
## 15        sum_ys_p100   2656.000      NA       NA         NA         NA
## 16      ratio_ys_mean      0.250   0.000    0.000       0.250       0.250
## 17        ratio_ys_sd      0.005   0.000    0.000       0.005       0.005
## 18    ratio_ys_median      0.250   0.000    0.000       0.250       0.250
## 19       ratio_ys_p25      0.247   0.000    0.000       0.246       0.247
## 20       ratio_ys_p75      0.253   0.000    0.000       0.253       0.253
## 21        ratio_ys_p0      0.218      NA       NA         NA         NA
## 22      ratio_ys_p100      0.306      NA       NA         NA         NA
## 23    prob_conclusive      0.669   0.005    0.005       0.659       0.678
## 24      prob_superior      0.053   0.002    0.002       0.048       0.057
## 25   prob_equivalence      0.616   0.005    0.005       0.607       0.626
```

```
## 26          prob_futility    0.000    0.000    0.000    0.000    0.000
## 27              prob_max     0.331    0.005    0.005    0.322    0.340
## 28 prob_select_arm_Arm A     0.327    0.005    0.005    0.318    0.336
## 29 prob_select_arm_Arm B     0.338    0.005    0.005    0.329    0.347
## 30 prob_select_arm_Arm C     0.335    0.005    0.005    0.326    0.344
## 31      prob_select_none     0.000    0.000    0.000    0.000    0.000
## 32                  rmse     0.010    0.000    0.000    0.010    0.011
## 33               rmse_te        NA       NA       NA       NA       NA
## 34                   mae     0.006    0.000    0.000    0.006    0.006
## 35                mae_te        NA       NA       NA       NA       NA
## 36                   idp        NA       NA       NA       NA       NA
```

```r
write.csv2(
  primary_performance_uncalibrated,
  file = paste0(dir_out, "Performance primary sims uncalibrated.csv"),
  row.names = FALSE
)
```

**Calibration**

```r
primary_design_null_scenario_calibration <- calibrate_trial(
  trial_spec = primary_design_null_scenario,
  n_rep = 10000,
  base_seed = 4131, # Reproducibility
  # Target, search range, tolerance, and maximum number of iterations
  target = 0.05,
  search_range = c(0.9, 1),
  tol = 0.001,
  dir = -1, # Only tolerate values below target, i.e., 0.049 to 0.050
  iter_max = 25,
  path = paste0(dir_out, "Primary calibration.RDS") # Save/reload
)

# Print summary of calibration results
primary_design_null_scenario_calibration
```

```
## Trial calibration:
## * Result: calibration successful
## * Best x: 0.9904165
## * Best y: 0.0492
##
## Central settings:
## * Target: 0.05
## * Tolerance: 0.001 (at or below target, range: 0.049 to 0.05)
## * Search range: 0.9 to 1
## * Gaussian process controls:
## * - resolution: 5000
## * - kappa: 0.5
## * - pow: 1.95
## * - lengthscale: 1 (constant)
## * - x scaled: yes
## * Noisy: no
## * Narrowing: yes
##
## Calibration/simulation details:
## * Total evaluations: 7 (previous + grid + iterations)
## * Repetitions: 10000
```

```
## * Calibration time: 1.22 hours
## * Base random seed: 4131
##
## See 'help("calibrate_trial")' for details.
```

```
# Check if successful (should be TRUE)
primary_design_null_scenario_calibration$success
```

```
## [1] TRUE
```

```
# Extract simulations conducted using the best stopping rule
primary_design_null_scenario_calibration$best_sims
```

```
## Multiple simulation results: generic binomially distributed outcome trial
## * Undesirable outcome
## * Number of simulations: 10000
## * Number of simulations summarised: 10000 (all trials)
## * No common control arm
## * Selection strategy: no selection if no superior arm
## * Treatment effect compared to: no comparison
##
## Performance metrics (using posterior estimates from final analysis [all patients]):
## * Sample sizes: mean 7935.0 (SD: 2414.4) | median 8950.0 (IQR: 5700.0 to 10000.0) [r
ange: 700.0 to 10000.0]
## * Total summarised outcomes: mean 1983.2 (SD: 604.9) | median 2248.0 (IQR: 1424.0 to
2486.0) [range: 157.0 to 2656.0]
## * Total summarised outcome rates: mean 0.250 (SD: 0.005) | median 0.250 (IQR: 0.247
to 0.253) [range: 0.218 to 0.306]
## * Conclusive: 66.1%
## * Superiority: 4.9%
## * Equivalence: 61.2%
## * Futility: 0.0% [not assessed]
## * Inconclusive at max sample size: 33.9%
## * Selection probabilities: Arm A: 1.5% | Arm B: 1.6% | Arm C: 1.8% | None: 95.1%
## * RMSE / MAE: 0.03233 / 0.02218
## * RMSE / MAE treatment effect: not estimated / not estimated
## * Ideal design percentage: not estimable
##
## Simulation details:
## * Simulation time: 15 mins
## * Base random seed: 4131
## * Credible interval width: 95%
## * Number of posterior draws: 10000
## * Estimation method: posterior medians with MAD-SDs
```

```
# Extract calibrated trial specification
primary_design_null_scenario_calibration$best_trial_spec
```

```
## Trial specification: generic binomially distributed outcome trial
## * Undesirable outcome
## * No common control arm
## * Best arms: Arm A and Arm B and Arm C
##
## Arms, true outcomes, starting allocation probabilities
## and allocation probability limits (min/max_probs rescaled):
##   arms true_ys start_probs fixed_probs min_probs max_probs
## Arm A   0.25       0.333          NA      0.25        NA
## Arm B   0.25       0.333          NA      0.25        NA
```

```
##  Arm C     0.25       0.333          NA       0.25          NA
##
## Maximum sample size: 10000
## Maximum number of data looks: 39
## Planned data looks after:  500, 750, 1000, 1250, 1500, 1750, 2000, 2250, 2500, 2750,
3000, 3250, 3500, 3750, 4000, 4250, 4500, 4750, 5000, 5250, 5500, 5750, 6000, 6250, 650
0, 6750, 7000, 7250, 7500, 7750, 8000, 8250, 8500, 8750, 9000, 9250, 9500, 9750, 10000
patients have reached follow-up
## Number of patients randomised at each look:  700, 950, 1200, 1450, 1700, 1950, 2200,
2450, 2700, 2950, 3200, 3450, 3700, 3950, 4200, 4450, 4700, 4950, 5200, 5450, 5700, 595
0, 6200, 6450, 6700, 6950, 7200, 7450, 7700, 7950, 8200, 8450, 8700, 8950, 9200, 9450,
9700, 9950, 10000
##
## Superiority threshold: 0.99042 (all analyses)
## Inferiority threshold: 0.00958 (all analyses)
## Equivalence thresholds:
## 1, 1, 1, 1, 0.9, 0.9, 0.9, 0.9, 0.9, 0.9, 0.9, 0.9, 0.9, 0.9, 0.9, 0.9, 0.9, 0.9, 0.
9, 0.9, 0.9, 0.9, 0.9, 0.9, 0.9, 0.9, 0.9, 0.9, 0.9, 0.9, 0.9, 0.9, 0.9, 0.9, 0.9, 0.9,
0.9, 0.9, 0.9
## (no common control)
## Absolute equivalence difference: 0.025
## No futility threshold (not relevant - no common control)
## Soften power for all analyses: 0.5
```

```
# Extract stopping threshold for superiority (using the default functionality
# the stopping threshold for inferiority is 1 - this value)
primary_design_null_scenario_calibration$best_x
```

```
## [1] 0.9904165
```

*Round stopping rules after calibration and re-assess*

```
# Extract and round calibrated stopping rule for superiority ('best_x')
superiority_rounded <- round(primary_design_null_scenario_calibration$best_x, 4)

# Extract calibrated trial design specification ('best_trial_spec') and update
# to use rounded stopping rules (inferiority = 1 - superiority)
primary_design_null_scenario_calib <-
  primary_design_null_scenario_calibration$best_trial_spec
primary_design_null_scenario_calib$superiority <- superiority_rounded
primary_design_null_scenario_calib$inferiority <- 1 - superiority_rounded

# Run large number of simulations with updated trial design specification
primary_null_calibrated <- run_trials(
  primary_design_null_scenario_calib,
  n_rep = 100000,
  path = paste0(dir_out, "Primary sims calibrated.RDS"), # Save/reload
  base_seed = 4131 # Reproducibility
)

# Check performance metrics without calculating uncertainty measures (not
# necessary due to the large number of simulations)
primary_performance_calibrated_rounded <- check_performance(
  primary_null_calibrated,
  select_strategy = "best"
)
```

```
# Print and save
primary_performance_calibrated_rounded

##                      metric        est
## 1             n_summarised 100000.000
## 2                size_mean   7931.986
## 3                  size_sd   2399.509
## 4              size_median   8950.000
## 5                 size_p25   5700.000
## 6                 size_p75  10000.000
## 7                  size_p0    700.000
## 8                size_p100  10000.000
## 9              sum_ys_mean   1983.076
## 10               sum_ys_sd    601.851
## 11           sum_ys_median   2246.000
## 12              sum_ys_p25   1429.000
## 13              sum_ys_p75   2485.000
## 14               sum_ys_p0    141.000
## 15             sum_ys_p100   2684.000
## 16           ratio_ys_mean      0.250
## 17             ratio_ys_sd      0.005
## 18         ratio_ys_median      0.250
## 19            ratio_ys_p25      0.247
## 20            ratio_ys_p75      0.253
## 21             ratio_ys_p0      0.201
## 22           ratio_ys_p100      0.306
## 23          prob_conclusive      0.664
## 24           prob_superior      0.048
## 25        prob_equivalence      0.616
## 26           prob_futility      0.000
## 27                prob_max      0.336
## 28      prob_select_arm_Arm A      0.329
## 29      prob_select_arm_Arm B      0.338
## 30      prob_select_arm_Arm C      0.333
## 31        prob_select_none      0.000
## 32                    rmse      0.010
## 33                 rmse_te         NA
## 34                     mae      0.006
## 35                  mae_te         NA
## 36                     idp         NA

write.csv2(
  primary_performance_calibrated_rounded,
  file = paste0(dir_out, "Performance primary sims calibrated and rounded.csv"),
  row.names = FALSE
)
```

**Performance assessment under other clinical scenarios**

First, all relevant scenarios (combinations of effects) are defined. The event probabilities are identical in arm A across all scenarios, but the unique combinations of small and large effects in both directions are specified for the other two arms. Of note, for simulation purposes, it does not matter which arm is which when there is no common control arm, except if using an arm selection strategy for performance metric calculation for simulations not ending with superiority where a specific arm (or a specific order of arms from a supplied list) is chosen. As the best remaining arm is selected when calculating performance metrics in this example, duplicate 'identical' combinations of event probabilities across arms B and C are not necessary and removed in the code below:

```r
# Possible combinations considered, 3 arms, 3 effect sizes
# Always constant event probabilities in arm A, only unique combinations of
# event probabilities in arms B and C used
effects <- c(0, 0.025, -0.025, 0.05, -0.05)
scenarios <- expand.grid(B = effects, C = effects)
scenarios$A <- 0
scenarios <- scenarios[, c("A", "B", "C")] # Reorder

# Remove non-unique combinations
for (i in nrow(scenarios):2) {
  cur_B <- scenarios[i, "B"]
  cur_C <- scenarios[i, "C"]
  remove <- FALSE
  for (j in (i-1):2) {
    if (scenarios[j, "B"] == cur_C & scenarios[j, "C"] == cur_B) {
      remove <- TRUE
    }
  }
  if (remove) {
    scenarios <- scenarios[-i, ]
  }
}
rownames(scenarios) <- 1:nrow(scenarios)

# Function for rounding and formatting results
rnd_fmt <- function(x, n = 0, mult = 1, suffix = "", na = "-") {
  res <- paste0(format(round(x * mult, digits = n), nsmall = n), suffix)
  if (!is.null(na)) {
    res <- ifelse(is.na(x), na, res)
  }
  res
}

# Prepare data.frame for formatted key results
key_results <- data.frame(
  A = paste0(rnd_fmt(scenarios$A + 0.25, 1, 100, "%")),
  B = paste0(rnd_fmt(scenarios$B + 0.25, 1, 100, "%")),
  C = paste0(rnd_fmt(scenarios$C + 0.25, 1, 100, "%")),
  size = "",
  pr_concl = NA,
  pr_sup = NA,
  pr_equi = NA
)

# Run 10,000 simulations for each new scenario (re-use previously calculated
# results for the null scenario), output results for each scenario, and extract
# key results
for (i in 1:nrow(scenarios)) {
  # Current scenario settings
  cur_scenario_name <- paste0(
    "A 25.0 - B ", format(25 + scenarios$B[i] * 100, nsmall = 1),
    " - C ", format(25 + scenarios$C[i] * 100, nsmall = 1))
  cur_true_ys <- c(0.25, 0.25 + scenarios$B[i], 0.25 + scenarios$C[i])
  if (all(cur_true_ys == 0.25)) { # Re-use null scenario results
```

```
    cur_sims <- primary_null_calibrated
} else {
    # Specify trial design with 'new' event probabilities and otherwise same settings
    cur_spec <- setup_trial_binom(
        arms = c("Arm A", "Arm B", "Arm C"),
        control = NULL,
        true_ys = cur_true_ys,
        highest_is_best = FALSE,
        start_probs = c(1/3, 1/3, 1/3),
        fixed_probs = NULL,
        min_probs = c(0.25, 0.25, 0.25),
        rescale_probs = "limits",
        soften_power = 0.5,
        data_looks = seq(from = 500, to = 10000, by = 250),
        randomised_at_looks = c(seq(from = 700, to = 9950, by = 250), 10000),
        inferiority = 1 - superiority_rounded,
        superiority = superiority_rounded,
        equivalence_prob = ifelse(seq(from = 500, to = 10000, by = 250) < 1500, 1, 0.9),
        equivalence_diff = 0.025,
        n_draws = 10000
    )
    # Run simulations
    cur_sims <- run_trials(
        cur_spec,
        n_rep = 10000,
        path = paste0(dir_out, "Primary ", cur_scenario_name, ".RDS"), # Save/reload
        base_seed = 4131 + i # Reproducibility
    )
}
# Summarise results in list format (without uncertainty measures) and print
cur_res <- summary(cur_sims, select_strategy = "best")
cat("\n\n###########################################",
    "\nPerformance metrics for scenario:", cur_scenario_name, "\n")
print(cur_res)
# Extract and save key results
key_results$size[i] <- rnd_fmt(cur_res$size_mean)
key_results$pr_concl[i] <- rnd_fmt(cur_res$prob_conclusive, 1, 100, "%")
key_results$pr_sup[i] <- rnd_fmt(cur_res$prob_superior, 1, 100, "%")
key_results$pr_equi[i] <- rnd_fmt(cur_res$prob_equivalence, 1, 100, "%")
}

## 
## 
## ###########################################
## Performance metrics for scenario: A 25.0 - B 25.0 - C 25.0
## Multiple simulation results: generic binomially distributed outcome trial
## * Undesirable outcome
## * Number of simulations: 1e+05
## * Number of simulations summarised: 1e+05 (all trials)
## * No common control arm
## * Selection strategy: best remaining available
## * Treatment effect compared to: no comparison
## 
## Performance metrics (using posterior estimates from final analysis [all patients]):
## * Sample sizes: mean 7932.0 (SD: 2399.5) | median 8950.0 (IQR: 5700.0 to 10000.0) [r
ange: 700.0 to 10000.0]
## * Total summarised outcomes: mean 1983.1 (SD: 601.9) | median 2246.0 (IQR: 1429.0 to
```

```
2485.0) [range: 141.0 to 2684.0]
## * Total summarised outcome rates: mean 0.250 (SD: 0.005) | median 0.250 (IQR: 0.247
to 0.253) [range: 0.201 to 0.306]
## * Conclusive: 66.4%
## * Superiority: 4.8%
## * Equivalence: 61.6%
## * Futility: 0.0% [not assessed]
## * Inconclusive at max sample size: 33.6%
## * Selection probabilities: Arm A: 32.9% | Arm B: 33.8% | Arm C: 33.3% | None: 0.0%
## * RMSE / MAE: 0.01027 / 0.00572
## * RMSE / MAE treatment effect: not estimated / not estimated
## * Ideal design percentage: not estimable
##
## Simulation details:
## * Simulation time: 3.14 hours
## * Base random seed: 4131
## * Credible interval width: 95%
## * Number of posterior draws: 10000
## * Estimation method: posterior medians with MAD-SDs
##
##
## ###################################################
## Performance metrics for scenario: A 25.0 - B 27.5 - C 25.0
## Multiple simulation results: generic binomially distributed outcome trial
## * Undesirable outcome
## * Number of simulations: 10000
## * Number of simulations summarised: 10000 (all trials)
## * No common control arm
## * Selection strategy: best remaining available
## * Treatment effect compared to: no comparison
##
## Performance metrics (using posterior estimates from final analysis [all patients]):
## * Sample sizes: mean 6495.9 (SD: 2389.3) | median 6200.0 (IQR: 4700.0 to 8700.0) [ra
nge: 700.0 to 10000.0]
## * Total summarised outcomes: mean 1656.2 (SD: 615.9) | median 1557.0 (IQR: 1181.0 to
2217.2) [range: 145.0 to 2723.0]
## * Total summarised outcome rates: mean 0.255 (SD: 0.007) | median 0.255 (IQR: 0.251
to 0.259) [range: 0.207 to 0.293]
## * Conclusive: 85.2%
## * Superiority: 14.6%
## * Equivalence: 70.6%
## * Futility: 0.0% [not assessed]
## * Inconclusive at max sample size: 14.8%
## * Selection probabilities: Arm A: 50.3% | Arm B: 0.6% | Arm C: 49.1% | None: 0.0%
## * RMSE / MAE: 0.01106 / 0.00518
## * RMSE / MAE treatment effect: not estimated / not estimated
## * Ideal design percentage: 99.4%
##
## Simulation details:
## * Simulation time: 5.21 mins
## * Base random seed: 4133
## * Credible interval width: 95%
## * Number of posterior draws: 10000
## * Estimation method: posterior medians with MAD-SDs
##
##
## ###################################################
```

```
## Performance metrics for scenario: A 25.0 - B 22.5 - C 25.0
## Multiple simulation results: generic binomially distributed outcome trial
## * Undesirable outcome
## * Number of simulations: 10000
## * Number of simulations summarised: 10000 (all trials)
## * No common control arm
## * Selection strategy: best remaining available
## * Treatment effect compared to: no comparison
##
## Performance metrics (using posterior estimates from final analysis [all patients]):
## * Sample sizes: mean 6473.4 (SD: 2779.0) | median 6450.0 (IQR: 4200.0 to 9450.0) [ra
nge: 700.0 to 10000.0]
## * Total summarised outcomes: mean 1539.8 (SD: 662.5) | median 1529.0 (IQR: 1015.0 to
2229.0) [range: 144.0 to 2564.0]
## * Total summarised outcome rates: mean 0.238 (SD: 0.007) | median 0.238 (IQR: 0.234
to 0.242) [range: 0.198 to 0.283]
## * Conclusive: 81.5%
## * Superiority: 59.7%
## * Equivalence: 21.8%
## * Futility: 0.0% [not assessed]
## * Inconclusive at max sample size: 18.5%
## * Selection probabilities: Arm A: 2.6% | Arm B: 94.6% | Arm C: 2.8% | None: 0.0%
## * RMSE / MAE: 0.01210 / 0.00578
## * RMSE / MAE treatment effect: not estimated / not estimated
## * Ideal design percentage: 94.6%
##
## Simulation details:
## * Simulation time: 9.23 mins
## * Base random seed: 4134
## * Credible interval width: 95%
## * Number of posterior draws: 10000
## * Estimation method: posterior medians with MAD-SDs
##
##
## #################################################
## Performance metrics for scenario: A 25.0 - B 30.0 - C 25.0
## Multiple simulation results: generic binomially distributed outcome trial
## * Undesirable outcome
## * Number of simulations: 10000
## * Number of simulations summarised: 10000 (all trials)
## * No common control arm
## * Selection strategy: best remaining available
## * Treatment effect compared to: no comparison
##
## Performance metrics (using posterior estimates from final analysis [all patients]):
## * Sample sizes: mean 5304.2 (SD: 1936.0) | median 4950.0 (IQR: 4200.0 to 6200.0) [ra
nge: 700.0 to 10000.0]
## * Total summarised outcomes: mean 1353.0 (SD: 491.1) | median 1257.0 (IQR: 1071.0 to
1596.0) [range: 157.0 to 2675.0]
## * Total summarised outcome rates: mean 0.256 (SD: 0.008) | median 0.255 (IQR: 0.251
to 0.260) [range: 0.224 to 0.310]
## * Conclusive: 97.1%
## * Superiority: 14.1%
## * Equivalence: 83.0%
## * Futility: 0.0% [not assessed]
## * Inconclusive at max sample size: 2.9%
## * Selection probabilities: Arm A: 50.2% | Arm B: 0.0% | Arm C: 49.8% | None: 0.0%
```

```
## * RMSE / MAE: 0.01146 / 0.00536
## * RMSE / MAE treatment effect: not estimated / not estimated
## * Ideal design percentage: 100.0%
##
## Simulation details:
## * Simulation time: 7.05 mins
## * Base random seed: 4135
## * Credible interval width: 95%
## * Number of posterior draws: 10000
## * Estimation method: posterior medians with MAD-SDs
##
##
## ################################################
## Performance metrics for scenario: A 25.0 - B 20.0 - C 25.0
## Multiple simulation results: generic binomially distributed outcome trial
## * Undesirable outcome
## * Number of simulations: 10000
## * Number of simulations summarised: 10000 (all trials)
## * No common control arm
## * Selection strategy: best remaining available
## * Treatment effect compared to: no comparison
##
## Performance metrics (using posterior estimates from final analysis [all patients]):
## * Sample sizes: mean 2870.7 (SD: 1501.5) | median 2700.0 (IQR: 1700.0 to 3700.0) [ra
nge: 700.0 to 10000.0]
## * Total summarised outcomes: mean 646.9 (SD: 336.0) | median 590.0 (IQR: 390.0 to 84
3.0) [range: 134.0 to 2304.0]
## * Total summarised outcome rates: mean 0.226 (SD: 0.010) | median 0.226 (IQR: 0.220
to 0.232) [range: 0.188 to 0.294]
## * Conclusive: 100.0%
## * Superiority: 99.6%
## * Equivalence: 0.4%
## * Futility: 0.0% [not assessed]
## * Inconclusive at max sample size: 0.0%
## * Selection probabilities: Arm A: 0.1% | Arm B: 99.8% | Arm C: 0.1% | None: 0.0%
## * RMSE / MAE: 0.01447 / 0.00757
## * RMSE / MAE treatment effect: not estimated / not estimated
## * Ideal design percentage: 99.8%
##
## Simulation details:
## * Simulation time: 4.07 mins
## * Base random seed: 4136
## * Credible interval width: 95%
## * Number of posterior draws: 10000
## * Estimation method: posterior medians with MAD-SDs
##
##
## ################################################
## Performance metrics for scenario: A 25.0 - B 27.5 - C 27.5
## Multiple simulation results: generic binomially distributed outcome trial
## * Undesirable outcome
## * Number of simulations: 10000
## * Number of simulations summarised: 10000 (all trials)
## * No common control arm
## * Selection strategy: best remaining available
## * Treatment effect compared to: no comparison
##
```

```
## Performance metrics (using posterior estimates from final analysis [all patients]):
## * Sample sizes: mean 6710.1 (SD: 2793.2) | median 6700.0 (IQR: 4450.0 to 9950.0) [ra
nge: 700.0 to 10000.0]
## * Total summarised outcomes: mean 1764.7 (SD: 736.5) | median 1790.0 (IQR: 1193.0 to
2560.0) [range: 163.0 to 2811.0]
## * Total summarised outcome rates: mean 0.263 (SD: 0.007) | median 0.263 (IQR: 0.259
to 0.267) [range: 0.229 to 0.307]
## * Conclusive: 77.5%
## * Superiority: 56.5%
## * Equivalence: 21.1%
## * Futility: 0.0% [not assessed]
## * Inconclusive at max sample size: 22.5%
## * Selection probabilities: Arm A: 94.3% | Arm B: 2.7% | Arm C: 3.0% | None: 0.0%
## * RMSE / MAE: 0.01263 / 0.00595
## * RMSE / MAE treatment effect: not estimated / not estimated
## * Ideal design percentage: 94.3%
##
## Simulation details:
## * Simulation time: 10.2 mins
## * Base random seed: 4137
## * Credible interval width: 95%
## * Number of posterior draws: 10000
## * Estimation method: posterior medians with MAD-SDs
##
##
## ################################################
## Performance metrics for scenario: A 25.0 - B 22.5 - C 27.5
## Multiple simulation results: generic binomially distributed outcome trial
## * Undesirable outcome
## * Number of simulations: 10000
## * Number of simulations summarised: 10000 (all trials)
## * No common control arm
## * Selection strategy: best remaining available
## * Treatment effect compared to: no comparison
##
## Performance metrics (using posterior estimates from final analysis [all patients]):
## * Sample sizes: mean 5052.5 (SD: 2550.0) | median 4700.0 (IQR: 3200.0 to 6950.0) [ra
nge: 700.0 to 10000.0]
## * Total summarised outcomes: mean 1215.1 (SD: 606.9) | median 1141.0 (IQR: 750.0 to
1637.0) [range: 151.0 to 2657.0]
## * Total summarised outcome rates: mean 0.242 (SD: 0.008) | median 0.241 (IQR: 0.236
to 0.246) [range: 0.211 to 0.296]
## * Conclusive: 95.2%
## * Superiority: 74.3%
## * Equivalence: 20.9%
## * Futility: 0.0% [not assessed]
## * Inconclusive at max sample size: 4.8%
## * Selection probabilities: Arm A: 4.4% | Arm B: 95.6% | Arm C: 0.0% | None: 0.0%
## * RMSE / MAE: 0.01308 / 0.00635
## * RMSE / MAE treatment effect: not estimated / not estimated
## * Ideal design percentage: 97.8%
##
## Simulation details:
## * Simulation time: 6.93 mins
## * Base random seed: 4138
## * Credible interval width: 95%
## * Number of posterior draws: 10000
```

```
## * Estimation method: posterior medians with MAD-SDs
##
##
## ##################################################
## Performance metrics for scenario: A 25.0 - B 30.0 - C 27.5
## Multiple simulation results: generic binomially distributed outcome trial
## * Undesirable outcome
## * Number of simulations: 10000
## * Number of simulations summarised: 10000 (all trials)
## * No common control arm
## * Selection strategy: best remaining available
## * Treatment effect compared to: no comparison
##
## Performance metrics (using posterior estimates from final analysis [all patients]):
## * Sample sizes: mean 5286.6 (SD: 2634.3) | median 4950.0 (IQR: 3200.0 to 7200.0) [ra
nge: 700.0 to 10000.0]
## * Total summarised outcomes: mean 1403.0 (SD: 693.1) | median 1327.0 (IQR: 871.0 to
1906.0) [range: 167.0 to 2851.0]
## * Total summarised outcome rates: mean 0.266 (SD: 0.008) | median 0.266 (IQR: 0.261
to 0.271) [range: 0.237 to 0.326]
## * Conclusive: 93.1%
## * Superiority: 73.0%
## * Equivalence: 20.1%
## * Futility: 0.0% [not assessed]
## * Inconclusive at max sample size: 6.9%
## * Selection probabilities: Arm A: 95.0% | Arm B: 0.0% | Arm C: 5.0% | None: 0.0%
## * RMSE / MAE: 0.01390 / 0.00642
## * RMSE / MAE treatment effect: not estimated / not estimated
## * Ideal design percentage: 97.5%
##
## Simulation details:
## * Simulation time: 7.41 mins
## * Base random seed: 4139
## * Credible interval width: 95%
## * Number of posterior draws: 10000
## * Estimation method: posterior medians with MAD-SDs
##
##
## ##################################################
## Performance metrics for scenario: A 25.0 - B 20.0 - C 27.5
## Multiple simulation results: generic binomially distributed outcome trial
## * Undesirable outcome
## * Number of simulations: 10000
## * Number of simulations summarised: 10000 (all trials)
## * No common control arm
## * Selection strategy: best remaining available
## * Treatment effect compared to: no comparison
##
## Performance metrics (using posterior estimates from final analysis [all patients]):
## * Sample sizes: mean 2349.8 (SD: 1282.6) | median 2200.0 (IQR: 1450.0 to 2950.0) [ra
nge: 700.0 to 9700.0]
## * Total summarised outcomes: mean 539.8 (SD: 286.6) | median 478.0 (IQR: 332.0 to 68
9.0) [range: 133.0 to 2166.0]
## * Total summarised outcome rates: mean 0.232 (SD: 0.011) | median 0.231 (IQR: 0.224
to 0.238) [range: 0.189 to 0.296]
## * Conclusive: 100.0%
## * Superiority: 99.8%
```

```
## * Equivalence: 0.2%
## * Futility: 0.0% [not assessed]
## * Inconclusive at max sample size: 0.0%
## * Selection probabilities: Arm A: 0.1% | Arm B: 99.9% | Arm C: 0.0% | None: 0.0%
## * RMSE / MAE: 0.01581 / 0.00826
## * RMSE / MAE treatment effect: not estimated / not estimated
## * Ideal design percentage: 99.9%
##
## Simulation details:
## * Simulation time: 3.29 mins
## * Base random seed: 4140
## * Credible interval width: 95%
## * Number of posterior draws: 10000
## * Estimation method: posterior medians with MAD-SDs
##
##
## ################################################
## Performance metrics for scenario: A 25.0 - B 22.5 - C 22.5
## Multiple simulation results: generic binomially distributed outcome trial
## * Undesirable outcome
## * Number of simulations: 10000
## * Number of simulations summarised: 10000 (all trials)
## * No common control arm
## * Selection strategy: best remaining available
## * Treatment effect compared to: no comparison
##
## Performance metrics (using posterior estimates from final analysis [all patients]):
## * Sample sizes: mean 6239.5 (SD: 2386.3) | median 5700.0 (IQR: 4450.0 to 8200.0) [ra
nge: 700.0 to 10000.0]
## * Total summarised outcomes: mean 1434.7 (SD: 553.9) | median 1329.0 (IQR: 1001.0 to
1900.0) [range: 139.0 to 2429.0]
## * Total summarised outcome rates: mean 0.230 (SD: 0.006) | median 0.230 (IQR: 0.226
to 0.234) [range: 0.198 to 0.274]
## * Conclusive: 87.2%
## * Superiority: 13.7%
## * Equivalence: 73.6%
## * Futility: 0.0% [not assessed]
## * Inconclusive at max sample size: 12.8%
## * Selection probabilities: Arm A: 0.5% | Arm B: 49.8% | Arm C: 49.7% | None: 0.0%
## * RMSE / MAE: 0.01050 / 0.00502
## * RMSE / MAE treatment effect: not estimated / not estimated
## * Ideal design percentage: 99.5%
##
## Simulation details:
## * Simulation time: 9.35 mins
## * Base random seed: 4141
## * Credible interval width: 95%
## * Number of posterior draws: 10000
## * Estimation method: posterior medians with MAD-SDs
##
##
## ################################################
## Performance metrics for scenario: A 25.0 - B 30.0 - C 22.5
## Multiple simulation results: generic binomially distributed outcome trial
## * Undesirable outcome
## * Number of simulations: 10000
## * Number of simulations summarised: 10000 (all trials)
```

```
## * No common control arm
## * Selection strategy: best remaining available
## * Treatment effect compared to: no comparison
##
## Performance metrics (using posterior estimates from final analysis [all patients]):
## * Sample sizes: mean 4715.6 (SD: 2525.0) | median 4450.0 (IQR: 2700.0 to 6450.0) [ra
nge: 700.0 to 10000.0]
## * Total summarised outcomes: mean 1135.3 (SD: 595.9) | median 1062.0 (IQR: 666.0 to
1519.0) [range: 149.0 to 2523.0]
## * Total summarised outcome rates: mean 0.243 (SD: 0.010) | median 0.242 (IQR: 0.237
to 0.247) [range: 0.211 to 0.301]
## * Conclusive: 96.4%
## * Superiority: 74.6%
## * Equivalence: 21.8%
## * Futility: 0.0% [not assessed]
## * Inconclusive at max sample size: 3.6%
## * Selection probabilities: Arm A: 4.8% | Arm B: 0.0% | Arm C: 95.2% | None: 0.0%
## * RMSE / MAE: 0.01346 / 0.00651
## * RMSE / MAE treatment effect: not estimated / not estimated
## * Ideal design percentage: 98.4%
##
## Simulation details:
## * Simulation time: 6.16 mins
## * Base random seed: 4142
## * Credible interval width: 95%
## * Number of posterior draws: 10000
## * Estimation method: posterior medians with MAD-SDs
##
##
## ##################################################
## Performance metrics for scenario: A 25.0 - B 20.0 - C 22.5
## Multiple simulation results: generic binomially distributed outcome trial
## * Undesirable outcome
## * Number of simulations: 10000
## * Number of simulations summarised: 10000 (all trials)
## * No common control arm
## * Selection strategy: best remaining available
## * Treatment effect compared to: no comparison
##
## Performance metrics (using posterior estimates from final analysis [all patients]):
## * Sample sizes: mean 4788.2 (SD: 2427.8) | median 4450.0 (IQR: 2950.0 to 6450.0) [ra
nge: 700.0 to 10000.0]
## * Total summarised outcomes: mean 1032.2 (SD: 517.0) | median 965.0 (IQR: 641.0 to 1
373.2) [range: 134.0 to 2362.0]
## * Total summarised outcome rates: mean 0.217 (SD: 0.008) | median 0.216 (IQR: 0.211
to 0.221) [range: 0.182 to 0.261]
## * Conclusive: 96.7%
## * Superiority: 75.5%
## * Equivalence: 21.2%
## * Futility: 0.0% [not assessed]
## * Inconclusive at max sample size: 3.3%
## * Selection probabilities: Arm A: 0.0% | Arm B: 95.5% | Arm C: 4.5% | None: 0.0%
## * RMSE / MAE: 0.01255 / 0.00627
## * RMSE / MAE treatment effect: not estimated / not estimated
## * Ideal design percentage: 97.7%
##
## Simulation details:
```

```
## * Simulation time: 6.64 mins
## * Base random seed: 4143
## * Credible interval width: 95%
## * Number of posterior draws: 10000
## * Estimation method: posterior medians with MAD-SDs
##
##
## #################################################
## Performance metrics for scenario: A 25.0 - B 30.0 - C 30.0
## Multiple simulation results: generic binomially distributed outcome trial
## * Undesirable outcome
## * Number of simulations: 10000
## * Number of simulations summarised: 10000 (all trials)
## * No common control arm
## * Selection strategy: best remaining available
## * Treatment effect compared to: no comparison
##
## Performance metrics (using posterior estimates from final analysis [all patients]):
## * Sample sizes: mean 3204.8 (SD: 1754.4) | median 2950.0 (IQR: 1950.0 to 4200.0) [ra
nge: 700.0 to 10000.0]
## * Total summarised outcomes: mean 881.9 (SD: 480.4) | median 795.0 (IQR: 518.8 to 11
55.0) [range: 162.0 to 3062.0]
## * Total summarised outcome rates: mean 0.276 (SD: 0.010) | median 0.275 (IQR: 0.270
to 0.282) [range: 0.227 to 0.341]
## * Conclusive: 99.8%
## * Superiority: 99.3%
## * Equivalence: 0.5%
## * Futility: 0.0% [not assessed]
## * Inconclusive at max sample size: 0.2%
## * Selection probabilities: Arm A: 99.8% | Arm B: 0.1% | Arm C: 0.1% | None: 0.0%
## * RMSE / MAE: 0.01544 / 0.00775
## * RMSE / MAE treatment effect: not estimated / not estimated
## * Ideal design percentage: 99.8%
##
## Simulation details:
## * Simulation time: 4.59 mins
## * Base random seed: 4144
## * Credible interval width: 95%
## * Number of posterior draws: 10000
## * Estimation method: posterior medians with MAD-SDs
##
##
## #################################################
## Performance metrics for scenario: A 25.0 - B 20.0 - C 30.0
## Multiple simulation results: generic binomially distributed outcome trial
## * Undesirable outcome
## * Number of simulations: 10000
## * Number of simulations summarised: 10000 (all trials)
## * No common control arm
## * Selection strategy: best remaining available
## * Treatment effect compared to: no comparison
##
## Performance metrics (using posterior estimates from final analysis [all patients]):
## * Sample sizes: mean 2211.2 (SD: 1263.4) | median 1950.0 (IQR: 1200.0 to 2950.0) [ra
nge: 700.0 to 10000.0]
## * Total summarised outcomes: mean 512.3 (SD: 279.7) | median 449.0 (IQR: 299.0 to 65
1.0) [range: 139.0 to 2168.0]
```

```
## * Total summarised outcome rates: mean 0.235 (SD: 0.013) | median 0.234 (IQR: 0.226
to 0.243) [range: 0.193 to 0.307]
## * Conclusive: 100.0%
## * Superiority: 99.8%
## * Equivalence: 0.2%
## * Futility: 0.0% [not assessed]
## * Inconclusive at max sample size: 0.0%
## * Selection probabilities: Arm A: 0.1% | Arm B: 99.9% | Arm C: 0.0% | None: 0.0%
## * RMSE / MAE: 0.01619 / 0.00870
## * RMSE / MAE treatment effect: not estimated / not estimated
## * Ideal design percentage: 100.0%
##
## Simulation details:
## * Simulation time: 2.99 mins
## * Base random seed: 4145
## * Credible interval width: 95%
## * Number of posterior draws: 10000
## * Estimation method: posterior medians with MAD-SDs
##
##
## ################################################
## Performance metrics for scenario: A 25.0 - B 20.0 - C 20.0
## Multiple simulation results: generic binomially distributed outcome trial
## * Undesirable outcome
## * Number of simulations: 10000
## * Number of simulations summarised: 10000 (all trials)
## * No common control arm
## * Selection strategy: best remaining available
## * Treatment effect compared to: no comparison
##
## Performance metrics (using posterior estimates from final analysis [all patients]):
## * Sample sizes: mean 4649.7 (SD: 1730.2) | median 4200.0 (IQR: 3700.0 to 5450.0) [ra
nge: 700.0 to 10000.0]
## * Total summarised outcomes: mean 954.9 (SD: 352.6) | median 883.0 (IQR: 755.0 to 11
18.0) [range: 125.0 to 2171.0]
## * Total summarised outcome rates: mean 0.206 (SD: 0.008) | median 0.205 (IQR: 0.201
to 0.210) [range: 0.179 to 0.257]
## * Conclusive: 99.0%
## * Superiority: 13.8%
## * Equivalence: 85.1%
## * Futility: 0.0% [not assessed]
## * Inconclusive at max sample size: 1.0%
## * Selection probabilities: Arm A: 0.0% | Arm B: 49.5% | Arm C: 50.4% | None: 0.0%
## * RMSE / MAE: 0.01096 / 0.00538
## * RMSE / MAE treatment effect: not estimated / not estimated
## * Ideal design percentage: 100.0%
##
## Simulation details:
## * Simulation time: 6.12 mins
## * Base random seed: 4146
## * Credible interval width: 95%
## * Number of posterior draws: 10000
## * Estimation method: posterior medians with MAD-SDs

# Print and save key results
key_results
```

```
##            A      B      C  size  pr_concl  pr_sup  pr_equi
## 1  25.0%  25.0%  25.0%  7932     66.4%    4.8%    61.6%
## 2  25.0%  27.5%  25.0%  6496     85.2%   14.6%    70.6%
## 3  25.0%  22.5%  25.0%  6473     81.5%   59.7%    21.8%
## 4  25.0%  30.0%  25.0%  5304     97.1%   14.1%    83.0%
## 5  25.0%  20.0%  25.0%  2871    100.0%   99.6%     0.4%
## 6  25.0%  27.5%  27.5%  6710     77.5%   56.5%    21.1%
## 7  25.0%  22.5%  27.5%  5052     95.2%   74.3%    20.9%
## 8  25.0%  30.0%  27.5%  5287     93.1%   73.0%    20.1%
## 9  25.0%  20.0%  27.5%  2350    100.0%   99.8%     0.2%
## 10 25.0%  22.5%  22.5%  6239     87.2%   13.7%    73.6%
## 11 25.0%  30.0%  22.5%  4716     96.4%   74.6%    21.8%
## 12 25.0%  20.0%  22.5%  4788     96.7%   75.5%    21.2%
## 13 25.0%  30.0%  30.0%  3205     99.8%   99.3%     0.5%
## 14 25.0%  30.0%  30.0%  2211    100.0%   99.8%     0.2%
## 15 25.0%  20.0%  20.0%  4650     99.0%   13.8%    85.1%
```

```r
write.csv2(
  key_results,
  file = paste0(dir_out, "Performance primary calibrated 15 scenarios.csv"),
  row.names = FALSE
)
```

**Log date and session info**

Save date and R/package versions for reproducibility:

```r
date()
```

```
## [1] "Wed Jan 15 11:47:59 2025"
```

```r
sessionInfo()
```

```
## R version 4.4.1 (2024-06-14 ucrt)
## Platform: x86_64-w64-mingw32/x64
## Running under: Windows 10 x64 (build 19045)
##
## Matrix products: default
##
##
## locale:
## [1] LC_COLLATE=Danish_Denmark.utf8  LC_CTYPE=Danish_Denmark.utf8
## [3] LC_MONETARY=Danish_Denmark.utf8 LC_NUMERIC=C
## [5] LC_TIME=Danish_Denmark.utf8
##
## time zone: Europe/Copenhagen
## tzcode source: internal
##
## attached base packages:
## [1] stats     graphics  grDevices utils     datasets  methods   base
##
## other attached packages:
## [1] adaptr_1.4.0
##
## loaded via a namespace (and not attached):
##  [1] compiler_4.4.1    fastmap_1.2.0     cli_3.6.3         parallel_4.4.1
```

```
##  [5] tools_4.4.1      htmltools_0.5.8.1 rstudioapi_0.17.1 yaml_2.3.10
##  [9] rmarkdown_2.29   knitr_1.49       xfun_0.50         digest_0.6.37
## [13] rlang_1.1.4      evaluate_1.0.1
```

# Appendix 2

This supplementary appendix includes supplementary examples of various trial designs along with explanations.

For additional details and examples, see the primary manuscript and the ***adaptr*** package documentation available at: https://inceptdk.github.io/adaptr/.

**Setup**

```
# Load adaptr
library(adaptr)

## Loading 'adaptr' package v1.4.0.
## For instructions, type 'help("adaptr")'
## or see https://inceptdk.github.io/adaptr/.

# Load ggplot2 (used for illustrations in this supplement)
library(ggplot2)
```

**Example 1: Design using a common control arm**

In this example, we specify a trial design specification with four arms with one arm (`'Standard'`) used as a common control arm to which the other arms are compared pairwise. If one of the other arms is superior to the initial common control in an adaptive analysis, the initial common control arm will be dropped and the arm deemed superior to this arm will become the new common control.

The design uses the special argument `control_prob_fixed`, which is set to `'sqrt-based'`. This means that the trial will use a fixed allocation probability to the common control arm, which is defined as the square root to the number of currently active non-control arms to 1 (for each active non-control arm). This will similarly be used for 'new' common control arms.

All non-control arms use response-adaptive randomisation, limited at minimum 15% (re-scaled proportionately when arms are dropped; not specified for the control arm, as this uses a fixed allocation probability) and with a *softening factor* of 0.5 applied.

The trial design specifies stopping rules for inferiority/superiority, but these could be calibrated if desired (as described in the main text). The trial design allows dropping non-control arms for practical equivalence if the probability that the absolute difference between a non-control arm and the current common control is less than 2.5%-points exceeds 90%. Similarly, non-control arms will be dropped for futility if the probability that they are NOT superior by at least 2.5%-points exceeds 90%. Practical equivalence and futility will only be assessed in comparisons to the initial common control.

In the scenario here, outcome distributions are identical in all arms (i.e., this represents a *null* scenario).

The `setup_trial_binom()` function is used.

```
design_common_control_null_scenario <- setup_trial_binom(
  # Arms and scenario
  arms = c("Standard", "Intervention A", "Intervention B", "Intervention C"),
  control = "Standard", # Common control arm
  true_ys = rep(0.25, 4),
  highest_is_best = FALSE,
  # Allocation rules
  min_probs = c(NA, 0.15, 0.15, 0.15),
  control_prob_fixed = "sqrt-based",
  rescale_probs = "limits",
  soften_power = 0.5,
  # participants with data/randomised at each analysis
  data_looks = seq(from = 500, to = 10000, by = 250),
  randomised_at_looks = pmin(seq(from = 500, to = 10000, by = 250) + 200, 10000),
  # stopping rules
  inferiority = 0.01, # default
  superiority = 0.99, # default
  equivalence_prob = 0.9,
  equivalence_diff = 0.025,
  equivalence_only_first = TRUE, # Only assessed against initial common control
  futility_prob = 0.9,
  futility_diff = 0.025,
  futility_only_first = TRUE, # Only assessed against initial common control
  # Posterior draws
  n_draws = 10000
)

# Print
design_common_control_null_scenario

## Trial specification: generic binomially distributed outcome trial
## * Undesirable outcome
## * Common control arm: Standard
## * Control arm probability fixed at 0.366 (for 4 arms), 0.414 (for 3 arms), 0.5 (for
## 2 arms)
## * Best arms: Standard and Intervention A and Intervention B and Intervention C
##
## Arms, true outcomes, starting allocation probabilities
## and allocation probability limits (min/max_probs rescaled):
##            arms true_ys start_probs fixed_probs min_probs max_probs
##        Standard    0.25       0.366       0.366        NA        NA
##   Intervention A    0.25       0.211          NA      0.15        NA
##   Intervention B    0.25       0.211          NA      0.15        NA
##   Intervention C    0.25       0.211          NA      0.15        NA
##
## Maximum sample size: 10000
## Maximum number of data looks: 39
## Planned data looks after:  500, 750, 1000, 1250, 1500, 1750, 2000, 2250, 2500, 2750,
## 3000, 3250, 3500, 3750, 4000, 4250, 4500, 4750, 5000, 5250, 5500, 5750, 6000, 6250, 650
## 0, 6750, 7000, 7250, 7500, 7750, 8000, 8250, 8500, 8750, 9000, 9250, 9500, 9750, 10000
## patients have reached follow-up
## Number of patients randomised at each look:  700, 950, 1200, 1450, 1700, 1950, 2200,
## 2450, 2700, 2950, 3200, 3450, 3700, 3950, 4200, 4450, 4700, 4950, 5200, 5450, 5700, 595
## 0, 6200, 6450, 6700, 6950, 7200, 7450, 7700, 7950, 8200, 8450, 8700, 8950, 9200, 9450,
## 9700, 9950, 10000
```

```
##
## Superiority threshold: 0.99 (all analyses)
## Inferiority threshold: 0.01 (all analyses)
## Equivalence threshold: 0.9 (all analyses) (only checked for first control)
## Absolute equivalence difference: 0.025
## Futility threshold: 0.9 (all analyses) (only checked for first control)
## Absolute futility difference (in beneficial direction): 0.025
## Soften power for all analyses: 0.5
```

**Example 2: Design using custom complex outcome-generating function**

*Example introduction*

This example specifies a trial design using the primary `setup_trial()` function, including custom outcome-generation and analysis functions, and explicit considerations regarding outcome data-lag and expected inclusion rates.

This example is inspired by an upcoming domain on the upcoming adaptive platform trial **INCEPT** (*The Intensive Care Platform Trial*, see INCEPT.dk). This will be a closed domain (i.e., arm adding is not permitted), and will thus to a large extent behave similar to a stand-alone advanced adaptive trial (see DOI: 10.1038/s41573-019-0034-3 for additional explanation).

As the outcome distribution considered is complex, the example is likewise relatively complex, to ensure that the outcome generation function provides a realistic outcome distribution.

The example uses two intervention arms, and the outcome considered here is ***days alive without life support at day 30***: a beneficial count outcome with a range of 0-29 (as participants will be on life support at inclusion). This outcome typically has substantial zero-inflation, which is considered when the outcome generation function is specified. Additional details on this and similar outcomes can be found elsewhere (DOI: 10.1186/s12874-023-01963-z).

*Distribution description and specification*

Here, data are generated from a 'hurdle-beta' two-part distribution (often referred to as a zero-inflated beta distribution, which is technically less correct) consisting of two sub-distributions. First, a binomial distribution controls the proportion of 0'es, which is set to 35% in this example. Second, a beta distribution controls the number of days in those with >0 days. As beta distributions model proportions (without including 0'es or 1's, i.e., only values >0% and <100% can be generated using this distribution), and as we only want 0'es to be generated by the binomial sub-distribution, the proportions obtained from the beta distribution are multiplied by the maximum possible number of days (29) and rounded upwards to the nearest whole number. This ensures that the beta sub-distribution does not generate any 0'es, but is able to generate the maximum value (29). While this will increase the overall mean of the outcome distribution generated slightly compared to the intended overall mean, this is unlikely to have any substantial influence on how the generated distribution resembles the intended reference distribution, and is considered smaller than the uncertainty associated with defining how a complex distribution like this is expected to be in a new trial population.

We want the overall distribution to have a mean number of days of 14; meaning that the mean number of days in those with >0 days should be approximately `14 / (1 - 0.35) = 21.5`, corresponding to a proportion of days of approximately 74.3% in those with >0 days. The variance component of the beta distribution here is set to be 0.05, which is roughly similar to the variance used in the upcoming INCEPT domain that this example is based upon (this was derived by fitting a beta distribution to the distribution of this outcome in those with >0 days in a previous trial that was considered likely to have a similar outcome distribution as expected here).

Of note, we have defined the Beta distribution using the *mean and variance* parameterisation as this is the easiest to use and interpret in this context. The functions built into R use the *alpha and beta* parametrisation (called *shape 1 and shape2*, respectively, in R), and thus we convert the parameters (see, e.g., [Wikipedia](#) for additional explanation, including on the parametrisations and formulae for converting from one parametrisation to another):

Define Beta distribution alpha and beta parameters:

```r
beta_mean <- 0.743 # Mean in those with >0 days in proportion scale
beta_var <- 0.05

beta_alpha <- abs((beta_mean * (beta_var + beta_mean^2 - beta_mean)) / beta_var)
beta_beta <- abs(( (beta_var + beta_mean^2 - beta_mean) * (beta_mean - 1) ) / beta_var)

cat("Beta distribution paramaters:",
    "\n- alpha (shape1):", beta_alpha,
    "\n- beta (shape2):", beta_beta)

## Beta distribution paramaters:
## - alpha (shape1): 2.094532
## - beta (shape2): 0.7244881
```

For illustrative purposes, we sample 100,000 values from this distribution using the values described above and similar rounding. As seen, the overall mean is slightly larger than the 14 days specified above due to rounding; this is considered acceptable (and the resulting distribution used below).

```r
set.seed(4131) # Reproducibility
# Sample each participants' chance of 0 days
sample_0_days <- rbinom(n = 100000, size = 1, prob = 0.35)
# Sample each participants' number of days if >0 days
sample_1p_days <- rbeta(
  n = 100000,
  shape1 = beta_alpha,
  shape2 = beta_beta,
)
# Combine both distributions and multiply and round to number of days
sample_days <- ceiling((1 - sample_0_days) * sample_1p_days * 29)

# Plot
ggplot(data = data.frame(days = sample_days), aes(x = days)) +
  geom_histogram(bins = 30) +
  scale_x_continuous(name = "Days alive without life support at day 30",
                     breaks = 0:5 * 5, expand = c(0, 0)) +
  scale_y_continuous(name = NULL, labels = function(x) (scales::percent(x / 100000)),
                     limits = c(0, 0.4 * 100000), expand = c(0, 0)) +
  labs(subtitle = paste("Mean:", round(mean(sample_days), 1),
                        "- Median:", round(median(sample_days), 1),
                        "- Interquartile range:", round(quantile(sample_days, probs = 0
.25), 1),
                        "to", round(quantile(sample_days, probs = 0.75), 1), "days")) +
  theme_bw()
```

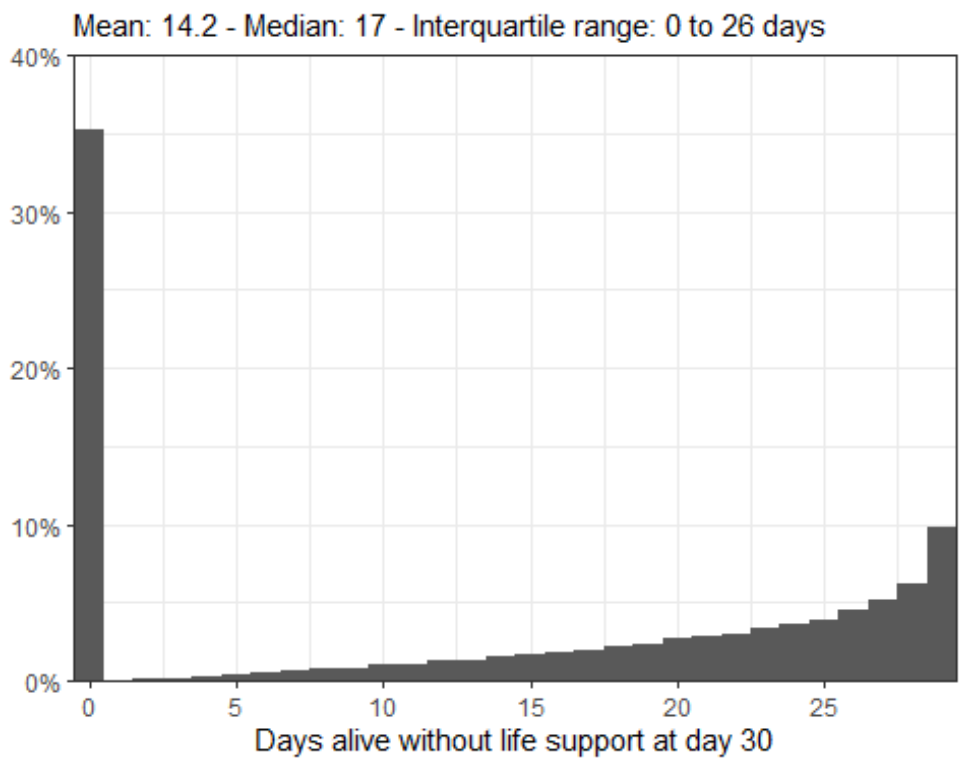

*Defining outcome-generating function*

Now, we specify the function to generate outcomes in the scenario with no differences between the two arms (denoted **'Arm A'** and **'Arm B'**), but in a format that makes this easy to change for scenarios with differences present. Instead of having one custom function for each scenario, advanced users may want to make a 'function factory' ([https://adv-r.hadley.nz/function-factories.html](https://adv-r.hadley.nz/function-factories.html)), but this is not done here to avoid overly complicating the example.

The requirements for outcome generation functions in `adaptr` are described in detail in the `setup_trial()` function documentation. Although outcome distributions areidentical in both arms in this scenario, we setup the function so that it loops through each arm, making it easy to change the most important distributional parameters, i.e., the proportion of zeroes and the proportion of days in those with > 0 days. Here, we choose to hardcode the beta distribution variance, as it is difficult to specify how this is expected to change between clinical scenarios. We thus generally consider it reasonable to assume that the variance parameter is unchanged between clinical scenarios, and that differences are mediated through changes to the proportions of zero days and/or the mean number of days for those with >0 days.

```r
# The only argument required is 'allocs', a character vector with all newly
# allocated participants that outcomes should be generated for
# The only return value should be a vector with outcomes (encoded as numerical
# values) for each participant in the same order as allocs
get_ys_hurdlebeta_days29_no_difference <- function(allocs) {
  # Setup empty return vector with outcomes
  y <- numeric(length(allocs))
  # Named vector of proportion of zeroes in each arm
  prop0 <- c("Arm A" = 0.35, "Arm B" = 0.35)
  # Named vector of the proportion of days in each arm in those with >0 days
  prop_days_1p <- c("Arm A" = 0.743, "Arm B" = 0.743)
  # Generate outcomes for each arm
  for (arm in c("Arm A", "Arm B")) {
```

```r
    ii <- which(allocs == arm) # Indices of participants in current arm
    y[ii] <- ceiling(
      (1 - rbinom(n = length(ii), size = 1, prob = prop0[arm])) * # Zeroes
        rbeta(n = length(ii),
              shape1 = abs((prop_days_1p[arm] * (0.05 + prop_days_1p[arm]^2 - prop_days
_1p[arm])) / 0.05),
              shape2 = abs(( (0.05 + prop_days_1p[arm]^2 - prop_days_1p[arm]) * (prop_d
ays_1p[arm] - 1) ) / 0.05)
        ) * 29) # Multiply and round
  }
  # Return outcomes
  y
}
```

Randomly sampling from the distribution using this function (with participants in both arms), generates similar distributions as the one illustrated above:

```r
set.seed(2024) # Reproducibility
ggplot(data = data.frame(
  days = get_ys_hurdlebeta_days29_no_difference(allocs = rep(c("Arm A", "Arm B"), 5000)
)),
  aes(x = days)) +
  geom_histogram(bins = 30) +
  scale_x_continuous(name = "Days alive without life support at day 30",
                     breaks = 0:5 * 5, expand = c(0, 0)) +
  scale_y_continuous(name = NULL, labels = function(x) (scales::percent(x / 10000)),
                     limits = c(0, 0.4 * 10000), expand = c(0, 0)) +
  labs(subtitle = paste("Mean:", round(mean(sample_days), 1),
                        "- Median:", round(median(sample_days), 1),
                        "- Interquartile range:", round(quantile(sample_days, probs = 0
.25), 1),
                        "to", round(quantile(sample_days, probs = 0.75), 1), "days")) +
  theme_bw()
```

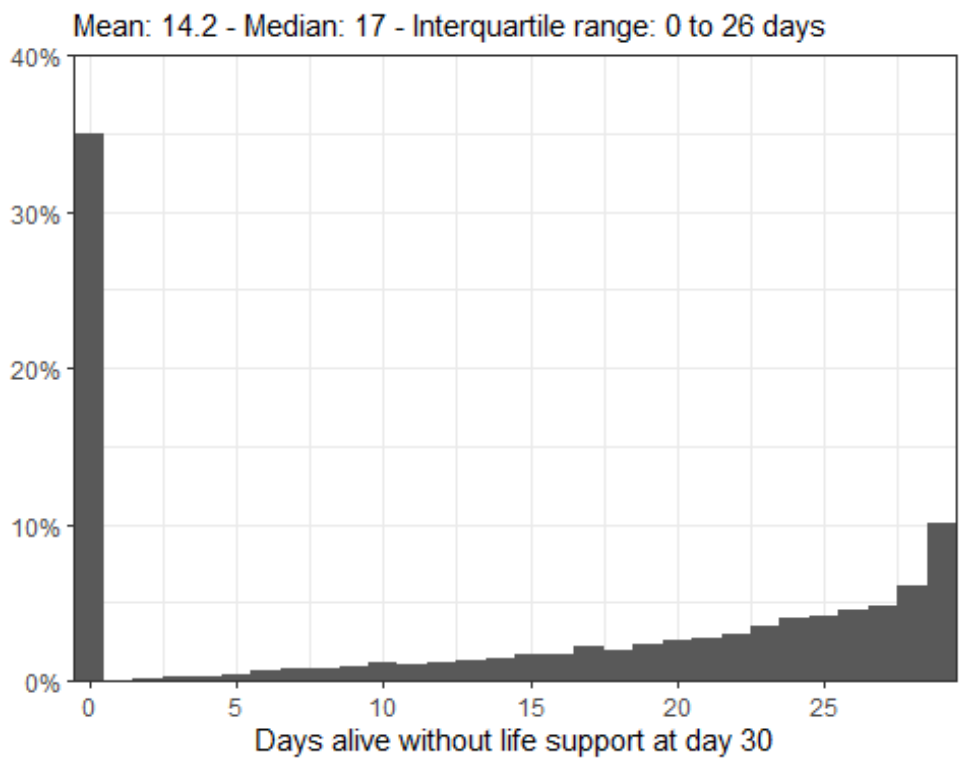

Of note, when using an outcome with a complex distribution like this and specifying scenarios with differences present, these may be specified to either only affect one part of the distribution (e.g., the proportion of 0'es or the means in those with values >0 in this example) or multiple parts, and, if so, to which extent differences are mediated by changes in the different parts of the distribution. It may be relevant to assess multiple sets of scenarios assessing differences mediated on different parts or combinations of parts of the distribution, ideally with one set of scenarios being the primary. Similarly, when doing sensitivity analyses of trial designs challenging the assumed outcome distributions, it may be relevant to assess the influence of similar overall changes in either direction but mediated via different parts or combinations of parts of the overall distribution.

*Defining analysis function*

In addition to the custom outcome-generating function, a custom function used to conduct the analysis and return the posterior draws is required when using the `setup_trial()` function. Here, we specify such a function in the correct format (additional details on the function arguments and required output is provided in the `setup_trial()` documentation).

Days alive without life support and similar outcomes may be analysed using different approaches, as discussed elsewhere ([DOI: 10.1186/s12874-023-01963-z](DOI: 10.1186/s12874-023-01963-z)).

In this example, we are interested in the overall means in each arm. In a 'real' trial, one option is to analyse this using linear regression, as modelling the entire distribution is of less interest than 'just' estimating the means in each arm and their differences. For simplicity, we thus simply generate posterior draws from normal distributions with means corresponding to the overall mean in each arm, and standard deviations defined as the the standard errors of the mean, using the conventional approach to calculation, i.e., the standard deviation in each arm divided by the square root to the number of participants in each arm minus 1. While this is an approximation that requires moderately large samples to provide adequately accurate measures of the standard errors of the means and thus the posterior distributions, we consider it adequate in this example as the first adaptive analysis is conducted after 500 participants have follow-up data available. In this example, we use no prior information in each arm, i.e., technically we use improper and completely flat priors. The function defined below corresponds to the simplified analysis function internally used by `setup_trial_norm()` in the current version of `adaptr`, just with added explanation:

```r
# The arguments that must be as specified here:
# - arms: character vector of all unique and currently active arms
# - allocs: character vector with allocations of all participants (in order of randomis
ation)
# - ys: character vector with outcomes of each participant (same order as allocs)
# - control: single character string, common control arm or NULL if none
# - n_draws: single integer, number of posterior draws in each arm
# Of note, even if no common control arm is used, the 'control' argument must be
# included as the other functions in the package supply this argument to all
# outcome-generating functions, as it may sometimes be relevant to use.
# The function must return a matrix of posterior draws (as numerical values)
# with one named column per currently active arm and with the number of rows
# corresponding to n_draws.
get_draws_norm_noprior <- function(arms, allocs, ys, control, n_draws) {
  draws <- list() # Prepare list to store results in
  for (arm in arms) { # Loop through all arms
    ii <- which(allocs == arm) # Indices of participants randomised to current arm
    n <- length(ii)
    if (n > 1){ # Return draws using the method described above if enough participants
randomised
      draws[[arm]] <- rnorm(n_draws, mean = mean(ys[ii]), sd = sd(ys[ii]) / sqrt(n - 1)
)
    } else {
      # Too few patients randomised - return extreme uncertainty based on the data
      # This is necessary to avoid errors if too few patients have been randomised to t
his arm yet
      draws[[arm]] <- rnorm(n_draws, mean = mean(ys), sd = 1000 * (max(ys) - min(ys)))
    }
  }
  do.call(cbind, draws) # Bind each vector contained in the list to a matrix
}
```

*Trial design specification*

We can now setup the complete trial specification using the functions defined above (these could also be directly specified within the `setup_trial()` call). Importantly, the `setup_trial()` function validates that the two custom functions run when supplied with the correct arguments and return outputs in the correct format, but it is the responsibility of the user to ensure that they otherwise work as intended (i.e., that the internal calculations are correct).

```r
design_complex_outcome_no_difference <- setup_trial(
  # Arm settings
  arms = c("Arm A", "Arm B"),
  control = NULL, # No common control
  # True overall means in each arm, using the means estimated from the combined
  # distribution after rounding (slightly higher than 14 days)
  true_ys = c(14.2, 14.2),
  # Custom functions to generate outcomes and posterior draws
  fun_y_gen = get_ys_hurdlebeta_days29_no_difference,
```

```
  fun_draws = get_draws_norm_noprior,
  highest_is_best = TRUE, # More days are better
  # Function to estimate the raw estimates (i.e., not from the posteriors) if
  # desired when calculating certain performance metrics - the default is to use
  # the posterior estimates in calculations, but this must always be specified
  fun_raw_est = mean, # Use the raw means
  # Use posterior medians (of the distribution of posterior means)for calculating performance metrics
  robust = TRUE,
  # Allocation rules
  start_probs = c(0.5, 0.5),, # Initial equal allocation
  fixed_probs = c(0.5, 0.5), # Fixed equal allocation
  # Participants with data available/randomised at each analysis
  data_looks = seq(from = 500, to = 10000, by = 250),
  randomised_at_looks = c(seq(from = 700, to = 9950, by = 250), 10000),
  # Stopping rules
  inferiority = 0.01,
  superiority = 0.99,
  equivalence_prob = ifelse(seq(from = 500, to = 10000, by = 250) < 1500, 1, 0.9),
  equivalence_diff = 0.025,
)

# Print
design_complex_outcome_no_difference

## Trial specification
## * Desirable outcome
## * No common control arm
## * Best arms: Arm A and Arm B
##
## Arms, true outcomes, starting allocation probabilities
## and allocation probability limits:
##   arms true_ys start_probs fixed_probs min_probs max_probs
## Arm A   14.2        0.5         0.5        NA        NA
## Arm B   14.2        0.5         0.5        NA        NA
##
## Maximum sample size: 10000
## Maximum number of data looks: 39
## Planned data looks after:  500, 750, 1000, 1250, 1500, 1750, 2000, 2250, 2500, 2750,
3000, 3250, 3500, 3750, 4000, 4250, 4500, 4750, 5000, 5250, 5500, 5750, 6000, 6250, 650
0, 6750, 7000, 7250, 7500, 7750, 8000, 8250, 8500, 8750, 9000, 9250, 9500, 9750, 10000
patients have reached follow-up
## Number of patients randomised at each look:  700, 950, 1200, 1450, 1700, 1950, 2200,
2450, 2700, 2950, 3200, 3450, 3700, 3950, 4200, 4450, 4700, 4950, 5200, 5450, 5700, 595
0, 6200, 6450, 6700, 6950, 7200, 7450, 7700, 7950, 8200, 8450, 8700, 8950, 9200, 9450,
9700, 9950, 10000
##
## Superiority threshold: 0.99 (all analyses)
## Inferiority threshold: 0.01 (all analyses)
## Equivalence thresholds:
## 1, 1, 1, 1, 0.9, 0.9, 0.9, 0.9, 0.9, 0.9, 0.9, 0.9, 0.9, 0.9, 0.9, 0.9, 0.9, 0.9, 0.
9, 0.9, 0.9, 0.9, 0.9, 0.9, 0.9, 0.9, 0.9, 0.9, 0.9, 0.9, 0.9, 0.9, 0.9, 0.9, 0.9, 0.9,
0.9, 0.9, 0.9
## (no common control)
## Absolute equivalence difference: 0.025
## No futility threshold (not relevant - no common control)
## Soften power for all analyses: 1 (no softening - all arms fixed)
```

**Example 3: Design using custom analysis function with custom priors**

*Example introduction*

Here, we specify a trial design using a custom function for generating posterior draws using informative priors. The example is inspired by the Empirical Meropenem versus Piperacillin/Tazobactam for Adult Patients with Sepsis (EMPRESS) trial ([DOI: 10.1111/aas.14441](DOI: 10.1111/aas.14441)) and a previous simulation study using `adaptr` ([DOI: 10.1002/pst.2387](DOI: 10.1002/pst.2387); code included in the supplement).

The example trial uses two arms and an undesirable binary outcome, (e.g., mortality). A custom outcome-generating function is specified (as this is required by `setup_trial()`), but this function is largely similar to the function used by `setup_trial_binom()`.

*Model and prior*

In practice, a trial like this could be analysed using a logistic regression model with a prior on the intervention effect (i.e., the difference between arms) specified on the log odds ratio scale. Here, we want to use an informative, neutral (centred on no difference) prior conveying scepticism towards large intervention effects. In the actual analysis, such a prior could be specified as a normal distribution with mean 0 and a standard deviation of, e.g., 0.5, corresponding to a distribution on the odds ratio scale centred on 1.00 and with 95% central probability mass between 0.38 and 2.66.

The amount of information this prior conveys can be 'translated' to the number of participants that would provide the same information in a trial with equal randomisation to both arms and a specific event rate in both groups. This calculation can be done according to a specific formula (see *Greenland S, Rothman KJ. Chapter 14: introduction to categorical statistics. In: KJ Rothman, TL Lash, S Greenland, eds. Modern Epidemiology. 3rd ed. Lippincott Williams & Wilkins; 2012: 237-257.* and [DOI: 10.1002/pst.2387](DOI: 10.1002/pst.2387)):

```
n = 1 / (prior_SD^2) * (4 / r + 4 / (1 - r) )
```

Where `n` is the total sample size (both arms combined), `prior_SD` is the standard deviation of the desired prior (normally distributed and with mean 0), and `r` is the event probability in both arms. Thus, a prior with a standard deviation of 0.5 corresponds to the same information as approximately 85 participants in a 'previous' trial with an event rate of 25% in both arms.

For simulation purposes, we use separate, conjugate beta-binomial models for each arm (see *Lambert B. Chapter 9: conjugate priors. In: B Lambert, ed. A Student's Guide to Bayesian Statistics. 1st ed. SAGE Publications Ltd.; 2018: 237-257.*). For these models, the prior can be specified as the number of participants with the outcome and without the outcome, respectively, in each arm. We can thus use the formula specified above to 'convert' the prior that will be used for the actual analyses (specified as a normally distributed prior for the difference between arms on the log odds ratio scale) to appropriate beta priors. First, we can derive the total sample size in a trial using the combined event rate estimate across both arms at any time and the formula above. Second, we then use half of the total estimated sample size for the beta prior in each arm, which then corresponds to this number of participants in total, with the event rate and non-event rate matching the combined current event/non-event rates in the current simulation. By doing so, we ensure that the priors have equal effect on the estimates in both arms, corresponding to the 'intended' prior in the actual planned analyses of the trial being neutral.

*Defining analysis function*

Here, we specify the function used to generate posterior draws in both arms according to the desired prior information and the details specified above. The required format, arguments, and outputs for functions generating posterior draws have been described in the previous example.

```r
fun_draws_prior <- function(arms, allocs, ys, control, n_draws) {
  # Total event rates and number of participants in a corresponding trial
  # with two arms, equal inclusion rates, and equal event rates
  prior_sd <- 0.5 # Prior standard deviation on log odds scale
  r <- mean(ys)
  n <- ( (1 / prior_sd^2) * ( ( 4 / r) + (4 / (1 - r)) ) )
  # In case of analyses with either no events or only events in all participants
  # n will be infinite regardless of the prior; in that case, n is set to 1,
  # just to ensure that simulations will not stop with an error if this occurs
  # due to chance in an early analysis
  if (!is.finite(n)) n <- 1
  # Prepare list to return posterior draws
  draws <- list()
  # Loop through all arms and generate posterior draws
  for (arm in arms) {
    ii <- which(allocs == arm) # Indices of current arm
    n_events <- sum(ys[ii]) # Total events in current arm
    draws[[arm]] <- rbeta(n_draws, # Posterior draws
                          (n * r / 2) + n_events,
                          (n * (1 - r) / 2) + length(ii) - n_events)
  }
  # Return draws after binding to matrix
  do.call(cbind, draws)
}
```

*Trial design specification*

Following this, the trial design is specified using the function defined above and a custom function for generating outcomes:

```r
design_custom_prior_no_difference <- setup_trial(
  # Arms and scenario
  arms = c("Arm A", "Arm B"),
  true_ys = c(0.25, 0.25),
  control = NULL,
  highest_is_best = FALSE,
  # Custom outcome generation function largely matching the default one used in
  # setup_trial_binom()
  fun_y_gen = function(allocs) {
    # Setup empty return vector with outcomes
    y <- rep(NA, length(allocs))
    # Named numeric vector with event probabilities in each arm (for easy changing)
    event_probs <- c("Arm A" = 0.25, "Arm B" = 0.25)
    # Generate outcomes for each arm
    for (arm in c("Arm A", "Arm B")) {
      ii <- which(allocs == arm) # Indices of participants in current arm
      y[ii] <- rbinom(n = length(ii), size = 1, prob = event_probs[arm])
    }
    # Return outcomes
```

```
    y
  },
  # Custom function to calculate and return posterior draws (defined above)
  fun_draws = fun_draws_prior,
  # Allocation rules (response-adaptive randomisation with minimum limits)
  start_probs = c(0.5, 0.5),
  min_probs = c(1/3, 1/3),
  # Participants with data available/randomised at each analysis
  # Lag of 350 participants, maximum values in both arguments should match, and
  # the length of both vectors must match (pmin truncates values >10000)
  data_looks = seq(from = 1000, to = 10000, by = 500),
  randomised_at_looks = pmin(seq(from = 1000, to = 10000, by = 500) + 350, 10000),
  # Stopping rules
  inferiority = 0.01,
  superiority = 0.99,
  equivalence_prob = 0.9,
  equivalence_diff = 0.025,
  # Posterior draws
  n_draws = 10000,
  robust = TRUE
)
```

## Log date and session info

Save date and R/package versions for reproducibility:

```
date()
```

```
## [1] "Wed Jan 15 12:12:56 2025"
```

```
sessionInfo()
```

```
## R version 4.4.1 (2024-06-14 ucrt)
## Platform: x86_64-w64-mingw32/x64
## Running under: Windows 10 x64 (build 19045)
##
## Matrix products: default
##
##
## locale:
## [1] LC_COLLATE=Danish_Denmark.utf8  LC_CTYPE=Danish_Denmark.utf8
## [3] LC_MONETARY=Danish_Denmark.utf8 LC_NUMERIC=C
## [5] LC_TIME=Danish_Denmark.utf8
##
## time zone: Europe/Copenhagen
## tzcode source: internal
##
## attached base packages:
## [1] stats     graphics  grDevices utils     datasets  methods   base
##
## other attached packages:
## [1] ggplot2_3.5.1 adaptr_1.4.0
##
## loaded via a namespace (and not attached):
## [1] vctrs_0.6.5      cli_3.6.3        knitr_1.49       rlang_1.1.4
## [5] xfun_0.50        generics_0.1.3   labeling_0.4.3   glue_1.8.0
```

```
##  [9] colorspace_2.1-1  htmltools_0.5.8.1  scales_1.3.0       rmarkdown_2.29
## [13] grid_4.4.1        evaluate_1.0.1     munsell_0.5.1      tibble_3.2.1
## [17] fastmap_1.2.0     yaml_2.3.10        lifecycle_1.0.4    compiler_4.4.1
## [21] dplyr_1.1.4       pkgconfig_2.0.3    rstudioapi_0.17.1  farver_2.1.2
## [25] digest_0.6.37     R6_2.5.1           tidyselect_1.2.1   pillar_1.10.1
## [29] parallel_4.4.1    magrittr_2.0.3     withr_3.0.2        tools_4.4.1
## [33] gtable_0.3.6
```